\documentclass[12pt]{article}
\pdfoutput=1
\usepackage{cite}
\usepackage[T1]{fontenc}
\usepackage[utf8]{inputenc}

\usepackage[english]{babel}

\usepackage{graphicx}
\usepackage[dvipsnames]{xcolor}

\usepackage{amsmath}
\usepackage{amsfonts}
\usepackage{amssymb}
\usepackage{amstext}
\usepackage{slashed}

\usepackage{multirow}
\usepackage{booktabs}

\usepackage[bookmarks, breaklinks, colorlinks,urlcolor=black, citecolor=red, 
linkcolor=blue]{hyperref}
\usepackage{placeins}
\usepackage{verbatim}

\usepackage{float} 

\def\dsp{\displaystyle}
\def\bc{\begin{center}}
\def\ec{\end{center}}
\def\be{\begin{equation}}
\def\ee{\end{equation}}
\def\beqn{\begin{eqnarray}}
\def\eeqn{\end{eqnarray}} 
\def\ec {\end{center}}
\def\nn {\nonumber}

\def\gev{\ensuremath{\mathrm{Ge\kern -0.1em V}}}
\def \Re{\text{Re}}
\def \Im{\text{Im}}

\def\YLQ{\ensuremath{Y_{\rm LQ}}}
\def\ZLQ{\ensuremath{Z_{\rm LQ}}}

\def\cL{{\cal L}}

\def\cM{{\cal M}}

\def\cP{{\cal P}}

\def\cO{{\cal O}}

\newcommand{\eqn}[1]{(\ref{#1})}
\newcommand{\bel}[1]{\be\label{#1}}

\def\tp#1{\textcolor{red}{#1}}

\begin{document}
	
	\renewcommand*{\thefootnote}{\fnsymbol{footnote}}
	
\mbox{}\hfill{IFIC/19-32}
\vskip 2cm	
	
	\begin{center}
		
{\Large\bf 
Constraints on scalar leptoquarks\\[10pt] from lepton and kaon physics} 
\\[10mm]
{Rusa Mandal$^{a,b}$\footnote{Email: Rusa.Mandal@uni-siegen.de}
and Antonio Pich$^a$}\footnote{Email: Antonio.Pich@ific.uv.es}
		
$^a$
{\small\em IFIC, Universitat de Val\`encia-CSIC, Apt. Correus 22085, E-46071 Val\`encia, Spain} \\[3mm]
$^b$
{\small\em Theoretische Elementarteilchenphysik, Naturwiss.- techn. Fakult$\ddot{a}$t, \\ Universit$\ddot{a}$t Siegen, 57068 Siegen, Germany}

	\end{center}

\begin{abstract}
We present a comprehensive analysis of low-energy signals of hypothetical scalar leptoquark interactions in lepton and kaon transitions. We derive the most general effective four-fermion Lagrangian induced by tree-level scalar leptoquark exchange and identify the Wilson coefficients predicted by the five possible types of scalar leptoquarks. The current constraints on the leptoquark Yukawa couplings arising from lepton and kaon processes are worked out, including also loop-induced transitions with only leptons (or quarks) as external states. In the presence of scalar leptoquark interactions, we also derive the differential distributions for flavour-changing neutral-current transitions in semileptonic kaon modes, including all known effects within the Standard Model. Their interference with the new physics contributions could play a significant role in future improvements of those constraints that are currently hampered by poorly-determined non-perturbative parameters.	
\end{abstract}

%

\section{Introduction}
\label{sec:Intro}

Many theories beyond the Standard Model (SM) which treat quarks and leptons in a similar footing include a particular type of bosons called `leptoquarks'. These particles are present in Gran Unified Theories, such as
$SU(5)$~\cite{Georgi:1974sy}, $SU(4)_C\times SU(2)_L\times SU(2)_R$~\cite{Pati:1974yy}, or $SO(10)$~\cite{Georgi:1974my}, where quarks and leptons usually appear in the same multiplets, but can also show up in some models with dynamical symmetry breaking like technicolor or extended technicolor \cite{Dimopoulos:1979es,Farhi:1980xs}.
Leptoquarks can turn a quark into a lepton and vice versa and, due to this unique nature, the discovery of leptoquarks would be an unambiguous signal of new physics (NP). 

Extensive searches for leptoquarks have been conducted in past experiments and the hunt is still very much on in recent colliders as well. 
So far, the LHC has not found any signal of new particles beyond the SM (BSM), pushing their mass scale further up to the TeV range. Under this circumstance, relatively low-energy phenomena can be important to indirectly identify the possible evidence of leptoquarks. At low energies, leptoquarks induce interactions between two leptons and two quarks, and/or four leptons (quarks), which are in some cases either stringently suppressed or forbidden in the SM. We use such measurements or upper bounds for decay modes from various experiments to constrain the leptoquark couplings. In most of the cases the analysis is done within the model-independent effective theory approach and thus can be used for other types of NP scenarios as well. 

There exist several thorough studies dealing with diverse aspects of leptoquark phenomenology~\cite{Buchmuller:1986iq,Davidson:1993qk,Hewett:1997ce,Dorsner:2016wpm,Arnold:2013cva}. Moreover, the experimental flavour anomalies recently observed in some $B$ decay modes have triggered a renewed interest in leptoquark interactions as a possible explanation of the data \cite{Dorsner:2013tla,Sakaki:2013bfa,Hiller:2014yaa,Bauer:2015knc,Alonso:2015sja,Freytsis:2015qca,Barbieri:2015yvd,Das:2016vkr,Becirevic:2016oho,Becirevic:2016yqi,Becirevic:2017jtw,Buttazzo:2017ixm,Choudhury:2017qyt,Bobeth:2017ecx,Bordone:2017bld,Blanke:2018sro,Greljo:2018tuh,Bordone:2018nbg,Becirevic:2018afm,Kumar:2018kmr,Angelescu:2018tyl,DiLuzio:2018zxy,Faber:2018qon,Mandal:2018kau,Fornal:2018dqn,Bar-Shalom:2018ure,Baker:2019sli,Cornella:2019hct,Hati:2019ufv,Gripaios:2014tna,Sahoo:2016pet,Crivellin:2017zlb,Cai:2017wry,deMedeirosVarzielas:2018bcy,Alok:2017jaf,Crivellin:2017dsk,Calibbi:2017qbu,Crivellin:2018yvo,Crivellin:2019qnh}.
In this paper we reconsider various decay modes which could be potential candidates to hint for possible evidence of leptoquark interactions. 
Since most of the recent analyses have focused on the leptoquark couplings to heavy quarks, we restrict ourselves to leptons and mesons with light quarks, namely kaons for this work. Many of the modes that we consider have been already studied in the past.
However, we carefully re-examine them by including almost all possible effects within the SM that were previously neglected. 
The experimental precision has been improved significantly in some cases and, therefore, the interference terms between the SM contribution and the NP interaction can be very important and need to be properly taken into account.

Leptoquarks can be scalar or vector particles. In this article we discuss only the scalar leptoquarks because they can be more easily analyzed in a model-independent way. The phenomenology of vector leptoquarks is much more sensitive to the ultra-violet (UV) completeness of a particular model. The particle content of the full UV theory can in principle affect the low-energy phenomena abruptly, hence the obtained limits on vector leptoquark couplings may not be that robust. Our bounds on the scalar leptoquark couplings are extracted from data at a low-energy scale of about few GeV. When constructing a full leptoquark theory, a proper scale evolution through renormalization group (RG) equations must be then incorporated.

The rest of the paper is organized in the following way. In Section~\ref{sec:Model} we briefly discuss a generic interaction Lagrangian, containing all five scalar leptoquarks, which will be the starting point of our analysis. The most general effective Lagrangian at low energies, arising from these scalar leptoquark interactions, is then derived in Section~\ref{sec:Eff}, where we also set up the notation. We discuss the bounds originating from several rare decays of leptons, and from their electric and magnetic moments, in Section~\ref{sec:Lepton}. The limits from kaon decays are derived in Section~\ref{sec:Kaon}.
Finally, in Section~\ref{sec:summary} we summarize our results.

\section{Scalar leptoquarks}
\label{sec:Model}

In this section we discuss the interaction of scalar leptoquarks with the SM fields. From the representations of quark and lepton fields under the SM gauge group $SU(3)_C\otimes SU(2)_L\otimes U(1)_Y$, the leptoquarks can be classified in five different categories. We follow the nomenclature widely used in the literature~\cite{Davidson:1993qk,Dorsner:2016wpm} for these leptoquarks as $S_1\,({\bf \bar{3},\,1},\,1/3)$, $\tilde{S}_1\,({\bf \bar{3},\,1},\,4/3)$, $R_2~({\bf 3,\,2},\,7/6)$, $\tilde{R}_2\,({\bf 3,\,2},\,1/6)$ and $S_3\,({\bf\bar{3},\,3},\,1/3)$. It can be seen that while $S_1$ and $\tilde{S_1}$ are singlets under the $SU(2)_L$ gauge group, $R_2$ and $\tilde{R_2}$ are doublets and $S_3$ transforms as a triplet.
The hypercharge normalization is chosen in such a way that the electromagnetic (EM) charge $Q_{\rm em}=T_3+Y$, where $T_3$ is the eigenvalue of the diagonal generator of $SU(2)_L$. 
We denote the left-handed SM quark (lepton) doublets as $Q$ ($L$), while $u_R$ ($d_R$) and $\ell_R$ are the right-handed up (down)-type quark and lepton singlets, respectively. 
The so-called genuine leptoquarks~\cite{Dorsner:2016wpm}, $R_2$ and $\tilde{R_2}$, can be assigned a definite baryon ($B=\frac{1}{3}$) and lepton ($L=-1$) number, but $S_1$, $\tilde{S_1}$ and $S_3$ could violate in principle the conservation of these quantum numbers through diquark interactions.
The leptoquark couplings to diquarks induce proton decays and thus have to be very suppressed. In this paper we neglect such couplings, as we are only interested in low-energy phenomena, and will assume that there is some symmetry in the UV theory that forbids these terms.

Using the freedom to rotate the different equal-charge fermion fields in flavour space, we adopt a `down' basis where the down-type-quark and charged-lepton Yukawas are diagonal. In this basis, the transformation from the fermion interaction eigenstates to mass eigenstates is simply given by
$u_L \to V^\dagger u_L$ and $\nu_L\to U \nu_L$, where $V$ is the quark Cabibbo–Kobayashi–Maskawa (CKM) matrix \cite{Cabibbo:1963yz,Kobayashi:1973fv} and $U$ the Pontecorvo–Maki–Nakagawa–Sakata (PMNS) unitary matrix in the neutrino sector \cite{Pontecorvo:1957cp,Maki:1962mu,Pontecorvo:1967fh}.

Following the notation used in Ref.~\cite{Dorsner:2016wpm}, we write the
fermionic interaction Lagrangian for the five mentioned scalar leptoquarks as
\begin{align}
\label{eq:Lfull}
\mathcal{L}_{\mathrm{LQ}}\; &=\; \overline{Q^c}\, i \tau_2\, Y_{\tiny S_1} L\; S_1 + \overline{ u^c_R} \,Z_{\tiny S_1} \ell_R\; S_1 \nn \\
&+ \overline{d^c_R}  \, Y_{\tiny \tilde{S_1}} \ell_R\; \tilde{S}_1 \nn \\
&+\overline \ell_R \,Y_{\tiny R_2}\, R_2^\dagger\, Q - \bar u_R \,Z_{\tiny R_2}\,
 R_2^T\, i \tau_2 \, L \nn \\
&- \overline{d}_R  \,Y_{\tiny \tilde{R_2}}\,  \tilde{R}_2^T\, i \tau_2 L\, \nn \\
& + \overline{Q^c}\, Y_{\tiny S_3} \,i\tau_2  \, {\boldsymbol \tau\bf \cdot S_3} \, L
+ \mathrm{h.c.} \, ,
\end{align}
where $f^c\equiv \mathcal{C}\bar f^{\, T}$ indicates the charge-conjugated field of the fermion $f$.
Here $Y_{\tiny\rm LQ}$ and $Z_{\tiny \rm LQ}$ are completely arbitrary Yukawa matrices in flavour space and $\tau_k,~k\in \{1,2,3\}$ are the Pauli matrices. We have suppressed the $SU(2)_L$ indices for simplicity.  Expanding the interaction terms in the mass-eigenstate basis we get
\begin{align}
\label{eq:Lfull2}
\mathcal{L}_{\mathrm{LQ}}\; &= \left[ {\overline{u^c_{L\!}}}^{\, i} (V^* Y_{\tiny S_1})^{ij}  \ell_L^j - {\overline{d^c_{L\!}}}^{\, i}\, (Y_{\tiny S_1}U)^{ij}\,  \nu_L^j + {\overline{ u^c_{R\!}}}^{\, i}\,  Z_{\tiny S_1}^{ij} \ell_R^j  \right] S_1 \nn \\
&+{\overline{d^c_{R\!}}}^{\, i}  \,Y_{\tiny \tilde{S_1}}^{ij}\, \ell_R^j\, \tilde{S}_1 \nn \\
&+\overline{\ell}_R^{\, i} Y_{\tiny R_2}^{ij} d_L^j\,{R_2^{2/3}}^* + \overline{ u}_R^{\, i} (Z_{\tiny R_2} U)^{ij}\, 
\nu_L^j\,R_2^{2/3}  +  \overline{\ell}_R^{\, i} (Y_{\tiny R_2} V^\dagger)^{ij}\,
u_L^j\,{R_2^{5/3}}^* - \overline{u}_R^{\, i} Z_{\tiny R_2}^{ij}  \ell_L^j\,R_2^{5/3}  \nn \\
& - \overline{d}_R^{\, i}  \,Y_{\tiny \tilde{R_2}}^{ij}\, \ell_L^j\, \tilde{R}_2^{2/3}  + \overline{d}_R^{\, i}  \,(Y_{\tiny \tilde{R_2}}U)^{ij}\, \nu_L^j\, \tilde{R}_2^{-1/3} \nn \\
&  -{\overline{d^c_{L\!}}}^{\, i}  (Y_{\tiny S_3}U)^{ij} \nu_L^j\, S_3^{1/3}  \!-\! \sqrt{2}\, {\overline{d^c_{L\!}}}^{\, i}  Y_{\tiny S_3}^{ij}\, \ell_L^j\, S_3^{4/3} + \sqrt{2}\, {\overline{u^c_{L\!}}}^{\, i}  (V^*Y_{\tiny S_3}U)^{ij} \nu_L^j\, S_3^{-2/3} - {\overline{u^c_{L\!}}}^{\, i}  (V^* Y_{\tiny S_3})^{ij} \ell_L^j\, S_3^{1/3} \nn \\
&+ \rm{h.c.}\,.
\end{align} 
We have explicitly shown the generation indices in Eq.~\eqref{eq:Lfull2}, where $i,j\in \{1,2,3\}$. The superscripts for $R_2,~\tilde{R_2}$ and $S_3$ denote the EM charge of the corresponding leptoquark. Being doublets, $R_2$ and $\tilde{R_2}$ each have two components, while for the triplet $S_3$ we get three components differing by their EM charges. As we neglect the diquark couplings, a baryon and lepton number can be assigned to all leptoquarks in a consistent way. In the subsequent sections we explore the constraints that arise on the arbitrary Yukawa matrices $Y_{\tiny\rm LQ}$ and $Z_{\tiny \rm LQ}$ from various lepton and kaon decay modes. 

\section{Effective Lagrangian }
\label{sec:Eff}

The tree-level exchange of scalar leptoquarks
($\phi = S_1,\, \tilde S_1, \, R_2,\, \tilde R_2, \, S_3$) generates a low-energy effective Lagrangian,

\bel{eq:L-LQ_0}
\cL_{\mathrm{eff}}^{\mathrm{LQ}}\, =\,
\cL_{\mathrm{eff}}^{\mathrm{cc}} + \cL_{\mathrm{eff}}^{\mathrm{nc},\ell}
+ \cL_{\mathrm{eff}}^{\mathrm{nc},\nu}\, ,
\ee
where ($i,k,m,n$ are flavour indices)
\beqn
\label{eq:Lcc}
\cL_{\mathrm{eff}}^{\mathrm{cc}}& =& \sum_{i,k,m,n}\left\{
[C_{V_L}]^{ik,mn}\, (\bar u_L^i\gamma^\mu d_L^k) (\bar \ell_L^m\gamma_\mu \nu_L^n) 
\, +\, 
[C_{V_R}]^{ik,mn}\, (\bar u_R^i\gamma^\mu d_R^k) (\bar \ell_L^m\gamma_\mu \nu_L^n)
\right.\nn\\ &&\left.\hskip .8cm\mbox{}\,
+\,  [C_{S_L}]^{ik,mn}\, (\bar u_R^i  d_L^k) (\bar \ell_R^m \nu_L^n)
\, +\, [C_{S_R}]^{ik,mn}\, (\bar u_L^i  d_R^k) (\bar \ell_R^m \nu_L^n)
\right.\nn\\[5pt] &&\left.\hskip .8cm\mbox{}\,
+\, [C_{T}]^{ik,mn}\, (\bar u_R^i\sigma^{\mu\nu} d_L^k) (\bar \ell_R^m\sigma_{\mu\nu} \nu_L^n)
\right\} \; +\; \mathrm{h.c.}\, ,
\eeqn
\beqn
\label{eq:LncLep}
\cL_{\mathrm{eff}}^{\mathrm{nc},\ell}&\!\!\! =&\!\!\! \sum_{i,k,m,n}\sum_{q=u,d}\left\{
[g_{V,q}^{LL}]^{ik,mn}\, (\bar q_L^i\gamma^\mu q_L^k) (\bar \ell_L^m\gamma_\mu \ell_L^n) 
\, +\, 
[g_{V,q}^{LR}]^{ik,mn}\, (\bar q_L^i\gamma^\mu q_L^k) (\bar \ell_R^m\gamma_\mu \ell_R^n) 
\right.\nn\\ &&\left.\hskip 1.5cm\mbox{}\, +\,
[g_{V,q}^{RL}]^{ik,mn}\, (\bar q_R^i\gamma^\mu q_R^k) (\bar \ell_L^m\gamma_\mu \ell_L^n) 
\, +\, 
[g_{V,q}^{RR}]^{ik,mn}\, (\bar q_R^i\gamma^\mu q_R^k) (\bar \ell_R^m\gamma_\mu \ell_R^n) 
\right.\nn\\[5pt] &&\left.\hskip 1.5cm\mbox{}\, +\,
[g_{S,q}^{L}]^{ik,mn}\, (\bar q_R^i q_L^k) (\bar \ell_R^m \ell_L^n) 
\, +\, 
[g_{S,q}^{R}]^{ik,mn}\, (\bar q_L^i q_R^k) (\bar \ell_L^m \ell_R^n) 
\right.\nn\\[5pt] &&\left.\hskip 1.5cm\mbox{}\, +\,
[g_{T,q}^{L}]^{ik,mn}\, (\bar q_R^i\sigma^{\mu\nu} q_L^k) (\bar \ell_R^m \sigma_{\mu\nu}\ell_L^n) \, +\,
[g_{T,q}^{R}]^{ik,mn}\, (\bar q_L^i\sigma^{\mu\nu} q_R^k) (\bar \ell_L^m \sigma_{\mu\nu}\ell_R^n) 
\right\}\, ,\quad
\eeqn
and
\be 
\label{eq:LncNeu}
\cL_{\mathrm{eff}}^{\mathrm{nc},\nu}\; = \; \sum_{i,k,m,n}\sum_{q=u,d}\left\{
[N_{V_L}^{q}]^{ik,mn}\, (\bar q_L^i\gamma^\mu q_L^k) (\bar \nu_L^m\gamma_\mu \nu_L^n) 
\, +\, 
[N_{V_R}^{q}]^{ik,mn}\, (\bar q_R^i\gamma^\mu q_R^k) (\bar \nu_L^m\gamma_\mu \nu_L^n) 
\right\}\, .
\ee

We detail next the contributions from the different scalar leptoquarks. Only those Wilson coefficients that are non-vanishing (up to Hermitian conjugation) are listed. Notice that the following coefficients do not receive any contribution from scalar leptoquark exchange:
\be 
C_{V_R} = C_{S_R} = g_{S,d}^{L} = g_{T,d}^{L} = g_{S,d}^{R} = g_{T,d}^{R} = 0\, .
\ee

\goodbreak
\noindent
\\{\large \bf $\bullet\quad \mathbf{S_1}$ exchange} 

All Wilson coefficients are proportional to $w_1\equiv 1/(2 M^2_{S_1})$. Defining $\{ C , g, N\} \equiv w_1\, \{ \hat C , \hat g, \hat N\}$, one gets:
\begin{align}
\label{eq:S1op}
[\hat C_{V_L}]^{ik,mn} & = - (Y_{S_1} U)^{kn}\; (V Y_{S_1}^*)^{im}\, ,
\qquad &
[\hat C_{S_L}]^{ik,mn} &= -4\, [\hat C_{T}]^{ik,mn} \, =\,
(Y_{S_1} U)^{kn}\; (Z_{S_1}^*)^{im}\, ,
\nn\\[2pt] 
[\hat g_{V,u}^{LL}]^{ik,mn} & = (V^* Y_{S_1})^{kn}\; (V Y_{S_1}^*)^{im}\, ,
& 
[\hat g_{S,u}^{L}]^{ik,mn} & = -4\, [\hat g_{T,u}^{L}]^{ik,mn}
\, =\, - (V^* Y_{S_1})^{kn}\; (Z_{S_1}^*)^{im}\, ,
\nn\\[2pt]
[\hat g_{V,u}^{RR}]^{ik,mn} & = (Z_{S_1})^{kn}\; (Z_{S_1}^*)^{im}\, ,
& 
[\hat g_{S,u}^{R}]^{ik,mn} & = -4\, [\hat g_{T,u}^{R}]^{ik,mn}
\, =\, - (Z_{S_1})^{kn}\; (V Y_{S_1}^*)^{im} \, ,
\nn\\[2pt] 
[\hat N_{V_L}^{d}]^{ik,mn} & = (Y_{S_1} U)^{kn}\; (Y_{S_1}^*U^*)^{im}\, .
& 
\end{align}

\noindent
\\{\large \bf $\bullet\quad \mathbf{\tilde{S}_1}$ exchange} 

Only one operator gets a non-zero contribution in this case:
\be 
\label{eq:S1Top}
[g_{V,d}^{RR}]^{ik,mn}\, =\, \frac{1}{2 M^2_{\tilde S_1}}\; Y_{\tilde S_1}^{kn} \, (Y_{\tilde S_1}^*)^{im}\, .
\ee

\noindent
\\{\large \bf $\bullet\quad \mathbf{R_2}$ exchange} 

Similar to the previous cases all Wilson coefficients are proportional to $w_2\equiv 1/(2 M^2_{R_2})$. Thus, we define $\{ C , g, N\} \equiv w_2\, \{ \hat C , \hat g, \hat N\}$. However, we separate the contributions from leptoquarks with different electric charges, so that leptoquark mass splittings can be easily taken into account. The exchange of $R_2^{2/3}$ gives
\begin{align}
\label{eq:R21op}
[\hat C_{S_L}]^{ik,mn} &= 4\, [\hat C_{T}]^{ik,mn} \, =\,
- (Z_{R_2} U)^{in}\; (Y_{R_2})^{mk}\, , &
\nn\\[2pt] 
[\hat g_{V,d}^{LR}]^{ik,mn} & = - (Y_{R_2}^\dagger)^{in}\; (Y_{R_2})^{mk}\, ,
& 
[\hat N_{V_R}^{u}]^{ik,mn} & = - (Z_{R_2} U)^{in}\; (Z_{R_2}^*U^* )^{km}\, ,
\end{align}
while $R_2^{5/3}$ exchange leads to
\begin{align}
\label{eq:R22op}
[\hat g_{V,u}^{LR}]^{ik,mn} & = - (V Y^\dagger_{R_2})^{in}\; (Y_{R_2} V^\dagger)^{mk}\, ,
& 
[\hat g_{S,u}^{L}]^{ik,mn} & = 4\, [\hat g_{T,u}^{L}]^{ik,mn}
\, =\,  (Z_{R_2})^{in}\; (Y_{R_2} V^\dagger)^{mk}\, ,
\nn\\[2pt]
[\hat g_{V,u}^{RL}]^{ik,mn} & = - (Z_{R_2})^{in}\; (Z_{R_2}^*)^{km}\, ,
& 
[\hat g_{S,u}^{R}]^{ik,mn} & = 4\, [\hat g_{T,u}^{R}]^{ik,mn}
\, =\, (V Y_{R_2}^\dagger)^{in}\; (Z_{R_2}^*)^{km} \, .
\end{align}

\noindent
\\{\large \bf $\bullet\quad \mathbf{\tilde{R}_2}$ exchange} 

The two different components of $\tilde{R}_2$ give rise to one operator each, given as
\be 
\label{eq:R2Top}
[g_{V,d}^{RL}]^{ik,mn}\, =\, - \frac{1}{2 M^2_{\tilde R_2^{2/3}}}\; Y_{\tilde R_2}^{in} \; (Y_{\tilde R_2}^*)^{km}\, ,
\qquad
[N_{V_R}^{d}]^{ik,mn}\, =\,
- \frac{1}{2 M^2_{\tilde R_2^{-1/3}}}\; (Y_{\tilde R_2}U)^{in} \; (Y_{\tilde R_2}^* U^*)^{km}\, .
\ee

\goodbreak
\noindent
\\{\large \bf $\bullet\quad \mathbf{S_3}$ exchange} 

All Wilson coefficients are proportional to $w_3\equiv 1/(2 M^2_{S_3})$. Again, we define $\{ C , g, N\} \equiv w_3\, \{ \hat C , \hat g, \hat N\}$, and separate the contributions from leptoquarks with different electric charges.  The exchange of $S_3^{1/3}$ gives
\begin{align}
\label{eq:S31op}
[\hat C_{V_L}]^{ik,mn} & = (V Y_{S_3}^*)^{im}\; (Y_{S_3} U)^{kn} \, ,
\qquad &
[\hat g_{V,u}^{LL}]^{ik,mn} & = (V Y_{S_3}^*)^{im}\; (V^* Y_{S_3})^{kn} \, ,
\nn \\[2pt]
[\hat N_{V_L}^{d}]^{ik,mn} & = (Y_{S_3}^*U^*)^{im}\; (Y_{S_3} U)^{kn} \, ,
&
\end{align}
while $S_3^{4/3}$ only contributes to
\be
\label{eq:S32op}
[\hat g_{V,d}^{LL}]^{ik,mn}  = 2\, (Y_{S_3}^*)^{im}\; (Y_{S_3})^{kn} \, ,
\ee
and $S_3^{-2/3}$ leads to
\be
\label{eq:S33op}
[\hat N_{V_L}^{u}]^{ik,mn} = 2\, (V Y_{S_3}^*U^*)^{im}\; (V^*Y_{S_3} U)^{kn} \, .
\ee

We denote the elements of the matrices $Y_{\rm LQ}$  and $Z_{\rm LQ}$  with lowercase, namely, $y_{\rm LQ}^{ij}$  and $z_{\rm LQ}^{ij}$. As we see from the above expressions, particular combinations of these Yukawa matrix elements arise several times, and hence for convenience we introduce the following notation:
\begin{eqnarray}
\label{eq:notation}
{y^\prime_{\rm LQ}}^{\!\!\!\!\!\!ij}\equiv (Y_{\rm LQ}\,V^\dagger)^{ij},
\qquad\qquad
{\tilde{y}_{\rm LQ}}^{ij}\equiv (V^*\, Y_{\rm LQ})^{ij},
\qquad\qquad
\hat{y}_{\rm LQ}^{ij}\equiv (Y_{\rm LQ}\,U)^{ij}.
\end{eqnarray}

\subsection{QCD running}

The previous derivation of the Wilson coefficients (matching calculation) applies at the high scale $\mu = M_{\mathrm{LQ}}$, where QCD interactions amount to very small corrections because $\alpha_s(M_{\mathrm{LQ}})$ is small. However, we need to evolve these predictions, using the renormalization group, to the much lower scales where hadronic decays take place. Neglecting electroweak corrections, we only need to care about the quark currents. One obtains then the simplified result:
\be 
W(\mu) \, =\, \Omega_W(\mu,M_{\mathrm{LQ}})\; W(M_{\mathrm{LQ}})\, ,
\ee
where W refers to any of the Wilson coefficients in Eqs.~\eqn{eq:Lcc} to \eqn{eq:LncNeu}. At lowest order (leading logarithm), the evolution operator is given by
\be 
\Omega_W(\mu,M_{\mathrm{LQ}})\, =\, \left(\frac{\alpha_s^{(n_f)}(\mu)}{\alpha_s^{(n_f)}(m_q^{f+1})}\right)^{-\gamma^J_1/\beta_1^{(n_f)}}
\cdots\quad
\left(\frac{\alpha_s^{(5)}(m_b)}{\alpha^{(5)}_s(m_t)}\right)^{-\gamma^J_1/\beta_1^{(5)}}\;
\left(\frac{\alpha_s^{(6)}(m_t)}{\alpha^{(6)}_s(M_{\mathrm{LQ}})}\right)^{-\gamma^J_1/\beta_1^{(6)}}
\ee
with $n_f$ the relevant number of quark flavours at the hadronic scale considered and $m_q^{f+1}$ the lightest (integrated out) heavy quark. The powers are governed by the first coefficients of the QCD $\beta$ function, $\beta_1^{(n_f)}= (2 n_f-33)/6$, and the current anomalous dimensions:
\be 
\gamma_1^V = 0\, ,\qquad\qquad
\gamma_1^S = 2\, ,\qquad\qquad
\gamma_1^T = -2/3\, .
\ee
Notice that the vector currents do not get renormalized, while the scalar and tensor currents renormalize multiplicatively. 
Electroweak corrections generate sizable mixings between the scalar and tensor operators \cite{Gonzalez-Alonso:2017iyc,Aebischer:2017gaw}.

\section{Bounds from leptons}
\label{sec:Lepton}
In this section we consider processes involving charged leptons in the initial and/or final states. This includes $\mu$ and $\tau$ decays, the electric and magnetic dipole moments of the electron and the muon, and $\mu$ conversion inside nuclei. These processes are either very precisely measured at experiments or very suppressed, and even in some cases they are disallowed in the SM. As a result, strong constraints can be imposed on the leptoquark couplings which induce such phenomena.

\subsection[$\tau $ decays to mesons]{$\boldsymbol{\tau}$ decays to mesons}
\label{subsec:tau}
The heaviest charged lepton in the SM, $\tau$, is the only lepton which can decay to mesons \cite{Pich:2013lsa}. Lepton-flavour-violating $\tau$ decays into mesons and a lighter  lepton $\ell = e, \mu$ are forbidden in the SM (up to tiny contributions proportional to neutrino masses that are completely negligible). However the leptoquark scenarios considered in this paper contribute to such decays at tree level. 
The experimental upper bounds on several $\tau^- \to P\,\ell^-$ and $\tau^- \to V \ell^-$ decay modes, where $P$ ($V$) is a pseudo-scalar (vector) meson,
put then strong constraints on the Yukawa couplings given in Eq.~\eqref{eq:Lfull2}.
After integrating out the leptoquarks, these decay modes get tree-level contributions from the neutral-current operators with charged leptons in Eq.~\eqref{eq:LncLep}~\cite{He:2019xxp,He:2019iqf,Babu:2019mfe}.

For a pseudoscalar final state with flavour quantum numbers $P_{ij}^0 \equiv q^i\bar q^j$, we find
\be 
\cM(\tau\to \ell\,P_{ij}^0)\, =\, \frac{i}{2}\, f_P\; \left\{ \alpha^{ij}_q\; (\bar\ell_L\tau_R) + \beta^{ij}_q\; (\bar\ell_R\tau_L)\right\}\, ,
\ee
where 
\beqn
\alpha^{ij}_q & = & \left[ m_\tau\, (g_{V,q}^{LL}-g_{V,q}^{RL}) - m_\ell\, (g_{V,q}^{LR}-g_{V,q}^{RR}) - \frac{m_P^2}{m_{q^i}+m_{q^j}}\, g_{S,q}^{R}
\right]^{ij,\ell 3}\, ,
\nn \\
\beta^{ij}_q & = & \left[ - m_\ell\, (g_{V,q}^{LL}-g_{V,q}^{RL}) + m_\tau\, (g_{V,q}^{LR}-g_{V,q}^{RR}) + \frac{m_P^2}{m_{q^i}+m_{q^j}}\, g_{S,q}^{L}
\right]^{ij,\ell 3}\, .
\eeqn
The QCD renormalization-scale dependence of the scalar Wilson coefficients $g_{S,q}^{R}$ and $g_{S,q}^{L}$ is exactly canceled by the running quark masses.

The numerical values of the meson decay constants $f_P$ are given in Appendix~\ref{app:FF}, where we compile the relevant hadronic matrix elements.
We remind that for the physical mesons one must take into account their quark structure. Thus,
\be 
\alpha_{\pi^0}\, =\, \frac{1}{\sqrt{2}}\, (\alpha^{11}_u-\alpha^{11}_d)\, ,
\qquad\qquad
\alpha_{K_{S,L}}\, \approx\, \frac{1}{\sqrt{2}}\, [\alpha^{12}_d (1 + \bar\epsilon_K)\mp \alpha^{21}_d (1 - \bar\epsilon_K)]\, ,
\ee
and similar expressions for $\beta_P$. 
The decay width is given by
\be 
\Gamma(\tau\to\ell\, P^0)\, =\, \frac{f_P^2\,\lambda_P^{1/2}}{128\pi m_\tau^3}\;\left\{
(m_\tau^2 + m_\ell^2 -m_P^2)\; \left( |\alpha_P|^2 + |\beta_P|^2\right)
+ 4 m_\tau m_\ell\,\mathrm{Re}\left(\alpha_P\beta^*_P\right)\right\} ,
\ee
with  $\lambda^{1/2}_{P,V}\equiv \lambda^{1/2}(m_\tau^2,m_\ell^2,m_{P,V}^2)$ where $\lambda(x,y,z)\equiv x^2+y^2+z^2-2xy-2xz-2yz$ is the 
K\"allen function.

\begin{table}[!t]
	
	\renewcommand{\arraystretch}{1.1}
	\resizebox{1.1\textwidth}{!}{\hspace*{-2.cm}
		\begin{tabular}{ c c c c c c c}
			\hline \hline
			& ${\rm BR_{exp}}$ &  \multicolumn{4}{c}{Scalar leptoquark couplings} & Bound \\
			\rm Mode &($\times10^{-8}$)  &$R_2$ & $S_1$ & $\tilde{R}_2,\,\tilde{S}_1$ & $S_3$ &$\times  \left(M_{\mathrm{LQ}}/{\rm TeV}\right)^4 $ \\[0.2ex] \hline
			\noalign{\vskip1pt}
			\multirow{4}{*}{$e \pi^0 $}  & \multirow{4}{*}{ $8.0$} & $|y_{13}^{\prime\,\dagger} y_{11}^\prime-y_{31}^*y_{11}|^2$ & $|\tilde{y}_{13}\tilde{y}^* _{11}|^2$ & $|y_{13} y_{11}^*|^2$ & $|\tilde{y}_{13}\tilde{y}^\dagger _{11}\!-\! 2\,y_{13}y_{11}^*|^2$& $6.0\times 10^{-4}$ \\[0.2ex]
			& & $|z_{13}z_{11}^*|^2$ & $|z_{13} z_{11}^*|^2$& &&$6.0\times 10^{-4}$ \\[0.2ex] 
			& & $|y_{13}^{\prime\,\dagger}z_{11}^*|^2$ & $|\tilde{y}_{13} z_{11}^*|^2$& &&$1.1\times 10^{-3}$ \\[0.2ex] 
			& & $|y_{11}^\prime z_{13}|^2$ & $|z_{13} \tilde{y}_{11}^*|^2$& &&$1.1\times 10^{-3}$ 
			\\[.2ex]\hline \noalign{\vskip 1pt}
			\multirow{4}{*}{$\mu \pi^0 $} &\multirow{4}{*}{$11.0$} & $|y_{13}^{\prime\,\dagger}y_{21}^\prime - y_{31}^*y_{21}|^2$ &$|\tilde{y}_{13}\tilde{y}^*_{12}|^2$ & $|y_{13} y_{12}^*|^2$  &$|\tilde{y}_{13}\tilde{y}^\dagger _{21}\!-\! 2\,y_{13}y_{12}^*|^2$& $8.3\times 10^{-4}$ \\[0.2ex]
			& & $|z_{13}z_{12}^*|^2$
			&$|z_{13} z_{12}^*|^2$ &  && $8.3\times 10^{-4}$ \\[0.2ex]
			& & $|y_{13}^{\prime\,\dagger}z_{12}^*|^2$
			&$|\tilde{y}_{13} z_{12}^*|^2$ & &&$1.5\times 10^{-3}$ \\[0.2ex]
			& & $|y_{21}^\prime z_{13}|^2$ &$|z_{13}\tilde{y}_{12}^*|^2$ & &&$1.5\times 10^{-3}$ 
			\\[.2ex]\hline \noalign{\vskip 1pt}
			$e K_S $    & $2.6 $ & $|y_{31}^*y_{12} - y_{32}^*y_{11}|^2$ & &$|y_{13}y_{21}^* - y_{23}y_{11}^*|^2$ &$4\, |y_{23}y_{11}^* - y_{13}y_{21}^*|^2$&$ 7.2\times 10^{-5}$ 
			\\[.2ex]\hline \noalign{\vskip 1pt}
			$\mu K_S $   & $2.3 $ & $|y_{31}^*y_{22}-y_{32}^*y_{21}|^2$ & & $|y_{13}y_{22}^* - y_{23}y_{12}^*|^2$ &$4\, |y_{23}y_{12}^* - y_{13}y_{22}^*|^2$&$ 6.4\times 10^{-5}$ 
			\\[.2ex]\hline \noalign{\vskip 1pt}
			\multirow{4}{*}{$e \eta $ } & \multirow{4}{*}{$9.2$} & $|y_{13}^{\prime\,\dagger} y_{11}^\prime+y_{31}^*y_{11}\!-\!1.7\, y_{32}^*y_{12}|^2$ & $|\tilde{y}_{13}\tilde{y}^*_{11}|^2$ & $|y_{13} y_{11}^*\!-\!1.7\,y_{23} y_{21}^*|^2$ & $|\tilde{y}_{13}\tilde{y}^*_{11}\!+\! 2\,y_{13}y_{11}^* \!-\!3.4\,y_{23}y_{21}^*|^2$ &$ 1.3\times 10^{-3}$ \\[0.2ex] 
			& &$|z_{13}z_{11}^*|^2$ & $|z_{13} z_{11}^*|^2$ & &&$ 1.3\times 10^{-3}$ \\[0.2ex] 
			& & $|y_{13}^{\prime\,\dagger}z_{11}^*|^2$ & $|\tilde{y}_{13} z_{11}^*|^2$& &&$ 8.0\times 10^{-6}$ \\[0.2ex] 
			& & $|y_{11}^\prime z_{13}|^2$ & $|z_{13} \tilde{y}_{11}^*|^2$& &&$ 8.0\times 10^{-6}$ 
			\\[.2ex]\hline \noalign{\vskip 1pt}
			\multirow{4}{*}{$\mu \eta$}  & \multirow{4}{*}{$6.5$} & $|y_{13}^{\prime\,\dagger} y_{21}^\prime+y_{31}^*y_{21}\!-\!1.7\, y_{32}^*y_{22}|^2$ & $|\tilde{y}_{13}\tilde{y}^*_{12}|^2$ & $|y_{13} y_{12}^*\!-\!1.7\,y_{23} y_{22}^*|^2$ & $|\tilde{y}_{13}\tilde{y}^*_{12}\!+\! 2\,y_{13}y_{12}^* \!-\!3.4\,y_{23}y_{22}^*|^2$ &$ 9.4\times 10^{-4}$ \\[0.2ex]
			& & $|z_{13}z_{12}^*|^2$
			&$|z_{13} z_{12}^*|^2$ &  &&$ 9.4\times 10^{-4}$ \\[0.2ex]
			& & $|y_{13}^{\prime\,\dagger}z_{12}^*|^2$
			&$|\tilde{y}_{13} z_{12}^*|^2$ & && $ 5.7\times 10^{-6}$\\[0.2ex]
			& & $|y_{21}^\prime z_{13}|^2$ &$|z_{13}\tilde{y}_{12}^*|^2$ & &&$ 5.7\times 10^{-6}$ 
			\\[.2ex]\hline \noalign{\vskip 1pt}
			\multirow{4}{*}{$e \eta^\prime $ } & \multirow{4}{*}{$16.0$} & $|y_{13}^{\prime\,\dagger} y_{11}^\prime+y_{31}^*y_{11}\!+\!2\, y_{32}^*y_{12}|^2$ & $|\tilde{y}_{13}\tilde{y}^*_{11}|^2$ & $|y_{13} y_{11}^*\!+\!2\,y_{23} y_{21}^*|^2$ & $|\tilde{y}_{13}\tilde{y}^*_{11}\!+\! 2\,y_{13}y_{11}^* \!+\!4\,y_{23}y_{21}^*|^2$ &$ 3.2\times 10^{-3}$ \\[0.2ex] 
			& &$|z_{13}z_{11}^*|^2$ & $|z_{13} z_{11}^*|^2$ & &&$ 3.2\times 10^{-3}$ \\[0.2ex] 
			& & $|y_{13}^{\prime\,\dagger}z_{11}^*|^2$ & $|\tilde{y}_{13} z_{11}^*|^2$& &&$ 3.0\times 10^{-6}$ \\[0.2ex] 
			& & $|y_{11}^\prime z_{13}|^2$ & $|z_{13} \tilde{y}_{11}^*|^2$& &&$ 3.0\times 10^{-6}$ 
			\\[.2ex]\hline \noalign{\vskip 1pt}
			\multirow{4}{*}{$\mu \eta^\prime$} & \multirow{4}{*}{$13.0 $} & $|y_{13}^{\prime\,\dagger} y_{21}^\prime+y_{31}^*y_{21}\!+\!2\, y_{32}^*y_{22}|^2$ & $|\tilde{y}_{13}\tilde{y}^*_{12}|^2$ & $|y_{13} y_{12}^*\!+\!2\,y_{23} y_{22}^*|^2$ & $|\tilde{y}_{13}\tilde{y}^*_{12}\!+\! 2\,y_{13}y_{12}^* \!+\!4\,y_{23}y_{22}^*|^2$ &$ 2.6\times 10^{-3}$ \\[0.2ex]
			& & $|z_{13}z_{12}^*|^2$
			&$|z_{13} z_{12}^*|^2$ &  &&$ 2.6\times 10^{-3}$ \\[0.2ex]
			& & $|y_{13}^{\prime\,\dagger}z_{12}^*|^2$
			&$|\tilde{y}_{13} z_{12}^*|^2$ & && $ 2.4\times 10^{-6}$\\[0.2ex]
			& & $|y_{21}^\prime z_{13}|^2$ &$|z_{13}\tilde{y}_{12}^*|^2$ & &&$ 2.4\times 10^{-6}$ \\[0.5ex]
			\hline
			\hline 
	\end{tabular}}
	\caption{90\% C.L. bounds on scalar leptoquark couplings from $\tau\to \ell\, P$ decays.}
	\label{Table:meson}
\end{table}

The decay amplitude into a vector final state takes the form
\beqn
\cM(\tau\to\ell\, V_{ij}^0)& =& \frac{1}{2}\, m_V f_V \varepsilon^\mu(k)^*\;
\left(\bar\ell\gamma_\mu \left[ (g_{V,q}^{LL}+g_{V,q}^{RL})\,\cP_L +
(g_{V,q}^{LR}+g_{V,q}^{RR})\,\cP_R\right]^{ij,\ell 3}\tau\right)
\nn\\ & - & 
i\, f_V^\perp(\mu)\, k^\mu\varepsilon^\nu(k)^*\; 
\left(\bar\ell\, \sigma_{\mu\nu} 
[g_{T,q}^{L}\cP_L + g_{T,q}^{R}\cP_R]^{ij,\ell 3}\tau\right)\, .
\eeqn
Denoting by $\alpha_V^L$, $\alpha_V^R$, $\alpha_V^{TL}$ and $\alpha_V^{TR}$
the corresponding combinations of leptoquark couplings 
$(g_{V,q}^{LL}+g_{V,q}^{RL})^{ij,\ell 3}$,
$(g_{V,q}^{LR}+g_{V,q}^{RR})^{ij,\ell 3}$, 
$(g_{T,q}^{L})^{ij,\ell 3}$ and 
$(g_{T,q}^{R})^{ij,\ell 3}$,
respectively, for a given vector meson $V$, we find,
\begin{align}
\Gamma(\tau\to\ell\, V^0)\, =&\, \frac{\lambda_V^{1/2}}{128\pi m_\tau^3}\;
\left[
f_V^2\,\left\{  [(m_\tau^2-m_\ell^2)^2+m_V^2\, (m_\tau^2+m_\ell^2) - 2 m_V^4]
\left( |\alpha_V^L|^2 + |\alpha_V^R|^2 \right)
\right.\right. \nn \\ & \hskip 2.7cm\left.\left.
-\, 12\, m_\tau m_\ell\, m_V^2\;\mathrm{Re}\left(\alpha_V^L \alpha_V^{R*}\right)\right\}
\right. \nn\\ & \left.
+\, 4 f_V^\perp(\mu)^2 \left\{ [2\, (m_\tau^2-m_\ell^2)^2-m_V^2\, (m_\tau^2+m_\ell^2) -  m_V^4] \left[ |\alpha_V^{TL}|^2 + |\alpha_V^{TR}|^2 \right]
\right.\right. \nn\\ & \hskip 2.7cm\left.\left.
-\, 12\, m_\tau m_\ell m_V^2\;\mathrm{Re}\left(\alpha_V^{TL} \alpha_V^{TR*}\right)\right\}
\right. \nn \\ & \left.
+\,12\, f_V f_V^\perp(\mu)\, m_V \left\{
m_\ell\, (m_\tau^2-m_\ell^2 + m_V^2)\;
\mathrm{Re}\left( \alpha_V^L \alpha_V^{TL*} + \alpha_V^R \alpha_V^{TR*}\right)
\right.\right. \nn\\ & \hskip 3.cm\left.\left.
-\, m_\tau\, (m_\tau^2-m_\ell^2 - m_V^2)\;
\mathrm{Re}\left( \alpha_V^L \alpha_V^{TR*} + \alpha_V^R \alpha_V^{TL*}\right)
\right\}
\right]\, .
\end{align}
The vector-meson couplings to the tensor quark currents, $f_V^\perp(\mu)$, are defined in Appendix~\ref{app:FF}, where their currently estimated values are also given.

\begin{table}[!t]
	
	\renewcommand{\arraystretch}{1.1}
	\resizebox{1.08\textwidth}{!}{\hspace*{-1.cm}
		\begin{tabular}{ c c c c c c c}
			\hline \hline
			& ${\rm BR_{exp}}$ &   \multicolumn{4}{c}{Scalar leptoquark couplings} & Bound \\
			\rm Mode &($\times10^{-8}$)  &$R_2$ & $S_1$ & $\tilde{R}_2,~(\tilde{S}_1)$ & $S_3$ &$\times  \left(M_{\rm LQ}/{\rm TeV}\right)^4 $ \\[0.2ex] \hline
			\noalign{\vskip1pt}
			\multirow{4}{*}{$e \rho^0$}   &\multirow{4}{*}{ $1.8$} & $|y_{13}^{\prime\,\dagger} y_{11}^\prime-y_{31}^*y_{11}|^2$ & $|\tilde{y}_{13}\tilde{y}^*_{11}|^2$ & $|y_{13} y_{11}^*|^2$  &$|\tilde{y}_{13}\tilde{y}^*_{11}- 2\,y_{13}y_{11}^*|^2$&$ 3.0\times 10^{-5}$ \\[0.2ex] 
			& & $|z_{13}z_{11}^*|^2$ & $|z_{13} z_{11}^*|^2$ &  &&$ 3.0\times 10^{-5}$ \\[0.2ex] 
			& & $|y_{13}^{\prime\,\dagger}z_{11}^*|^2$ & $|\tilde{y}_{13} z_{11}^*|^2$& &&$1.2\times 10^{-4}$ \\[0.2ex] 
			& & $|y_{11}^\prime z_{13}|^2$ & $|z_{13} \tilde{y}_{11}^*|^2$& &&$1.2\times 10^{-4}$ 
\\[.2ex]\hline \noalign{\vskip 1pt} 
			\multirow{4}{*}{$\mu \rho^0$} & \multirow{4}{*}{$1.2$} & $|y_{13}^{\prime\,\dagger}y_{21}^\prime - y_{31}^*y_{21}|^2$ &$|\tilde{y}_{13}\tilde{y}^*_{12}|^2$ & $|y_{13} y_{12}^*|^2$  &$|\tilde{y}_{13}\tilde{y}^*_{12}- 2\,y_{13}y_{12}^*|^2$&$ 2.0\times 10^{-5}$ \\[0.2ex]
			& & $|z_{13}z_{12}^*|^2$ &$|z_{13} z_{12}^*|^2$ &  && $2.0\times 10^{-5}$ \\[0.2ex]
			& & $|y_{13}^{\prime\,\dagger}z_{12}^*|^2$
			&$|\tilde{y}_{13} z_{12}^*|^2$ & &&$7.8\times 10^{-5}$ \\[0.2ex]
			& & $|y_{21}^\prime z_{13}|^2$ &$|z_{13}\tilde{y}_{12}^*|^2$ & &&$7.8\times 10^{-5}$ 
\\[.2ex]\hline \noalign{\vskip 1pt}
			\multirow{4}{*}{$e \omega $}  & \multirow{4}{*}{$4.8$} & $|y_{13}^{\prime\,\dagger} y_{11}^\prime +y_{31}^*y_{11}|^2$ & $|\tilde{y}_{13}\tilde{y}^*_{11}|^2$ & $|y_{13} y_{11}^*|^2$  &$|\tilde{y}_{13}\tilde{y}^*_{11}+ 2\,y_{13}y_{11}^*|^2$&$ 9.9\times 10^{-5}$ \\[0.2ex] 
			& & $|z_{13}z_{11}^*|^2$ & $|z_{13} z_{11}^*|^2$ & && $ 9.9\times 10^{-5}$\\[0.2ex]
			& & $|y_{13}^{\prime\,\dagger}z_{11}^*|^2$ & $|\tilde{y}_{13} z_{11}^*|^2$& &&$3.1\times 10^{-4}$ \\[0.2ex] 
			& & $|y_{11}^\prime z_{13}|^2$ & $|z_{13} \tilde{y}_{11}^*|^2$& &&$3.1\times 10^{-4}$ 
\\[.2ex]\hline \noalign{\vskip 1pt}
			\multirow{4}{*}{$\mu \omega$} & \multirow{4}{*}{$4.7$} & $|y_{13}^{\prime\,\dagger}y_{21}^\prime + y_{31}^*y_{21}|^2$ &$|\tilde{y}_{13}\tilde{y}^*_{12}|^2$ & $|y_{13} y_{12}^*|^2$ &$|\tilde{y}_{13}\tilde{y}^*_{12}+ 2\,y_{13}y_{12}^*|^2$& $ 9.8\times 10^{-5}$ \\[0.2ex]
			&  & $|z_{13}z_{12}^*|^2$ &$|z_{13} z_{12}^*|^2$ &  && $ 9.8\times 10^{-5}$\\[0.2ex]
			& & $|y_{13}^{\prime\,\dagger}z_{12}^*|^2$
			&$|\tilde{y}_{13} z_{12}^*|^2$ & &&$3.1\times 10^{-4}$ \\[0.2ex]
			& & $|y_{21}^\prime z_{13}|^2$ &$|z_{13}\tilde{y}_{12}^*|^2$ & &&$3.1\times 10^{-4}$ 
\\[.2ex]\hline \noalign{\vskip 1pt}
			$e K^{*0} $  & $3.2 $ & $|y_{31}^*y_{12}|^2$ & &
		$|y_{13}y_{21}^*|^2	,(|y_{23}y_{11}^*|^2)$ &
		$4\, |y_{23}y_{11}^*|^2$ &$ 5.8\times 10^{-5}$ 
\\[.2ex]\hline \noalign{\vskip 1pt}
			$\mu K^{*0}$ & $5.9$ & $|y_{31}^*y_{22}|^2$ & & 
			$|y_{13}y_{22}^*|^2 ,(|y_{23}y_{12}^*|^2)$ 
			&$4\, |y_{23}y_{12}^*|^2$ &$1.1\times 10^{-4}$ 
\\[.2ex]\hline \noalign{\vskip 1pt}
			$e \overline{K}^{*0} $  & $3.4 $ & $|y_{32}^*y_{11}|^2$ & &
			$|y_{23}y_{11}^*|^2,(|y_{13}y_{21}^*|^2)$ &
			$4\, |y_{13}y_{21}^*|^2$ &$ 6.2\times 10^{-5}$ 
\\[.2ex]\hline \noalign{\vskip 1pt}
			$\mu \overline{K}^{*0}$ & $7.0$ & $|y_{32}^*y_{21}|^2$ & & 
			$|y_{23}y_{12}^*|^2 ,(|y_{13}y_{22}^*|^2)$
			&$4\, |y_{13}y_{22}^*|^2$ &$1.3\times 10^{-4}$ 
\\[.2ex]\hline \noalign{\vskip 1pt}
			$e \phi $    & $3.1 $ & $|y_{32}^*y_{12}|^2$ & & $|y_{23}y_{21}^*|^2$ 
			& 4\,$|y_{23}y_{21}^*|^2$ &$5.1\times 10^{-5}$ 
\\[.2ex]\hline \noalign{\vskip 1pt}
			$\mu \phi$   & $8.4 $ & $|y_{32}^*y_{22}|^2$ & & $|y_{23}y_{22}^*|^2$
			& 4\,$|y_{23}y_{22}^*|^2$ &$1.4\times 10^{-4}$ \\[0.2ex]
			\hline
			\hline 
	\end{tabular}}
	\caption{90\% C.L. bounds on scalar leptoquark couplings from $\tau\to \ell\, V$ decays.}
\label{Table:meson2}
\end{table}

There are strong experimental (90\% C.L.) upper bounds on the $\tau\to \ell\, P$ and $\tau\to \ell\, V$ decay modes, with $P=\pi^0,K_S,\eta,\eta^\prime$ and $V=\rho^0,\omega,K^{*0},\overline{K}^{*0},\phi$  \cite{Tanabashi:2018oca}. In Tables~\ref{Table:meson} and \ref{Table:meson2} we highlight the corresponding upper limits on the product of leptoquark Yukawa couplings that arise from such measurements, for the five different types of scalar leptoquarks. 
Columns 3 to 6 indicate the combinations of couplings that get bounded in each case.
For simplicity we have dropped the subscript with the leptoquark name in the Yukawa matrix elements. 
The upper bounds on these couplings are given in the last column of the Tables.
Notice that the limits scale with $M_{\mathrm{LQ}}^4$ (the numbers correspond to $M_{\mathrm{LQ}}= 1~\mathrm{TeV}$) and deteriorate very fast with increasing leptoquark masses.

For the $R_2$ and $S_1$ leptoquarks,
the decay amplitudes $\tau\to \pi^0\ell,\eta\ell,\eta^\prime\ell$ and $\tau\to \rho^0 \ell,\omega\ell$
can receive contributions from several combinations of couplings that 
we have separated in four rows. The first two correspond to contributions from vector and axial-vector operators, which can arise when either $\YLQ$ or $\ZLQ$ is non-zero. The first row assumes  $\ZLQ=0$ in order to bound $\YLQ$, while the opposite is done in the second row.
The pseudoscalar and tensor operators can only generate contributions when both $\YLQ$ and $\ZLQ$ are non-vanishing; the corresponding combinations of couplings are given in the third and fourth rows, and their limits assume all other contributions to be absent. Obviously, these bounds are weaker since they neglect possible interference effects that could generate fine-tuned cancellations.

We have neglected the tiny $CP$-violating component of the $K_S$ state. We remind that the `prime' and `tilde' notations imply the inclusion of the CKM matrix $V$ as defined in Eq.~\eqref{eq:notation}. In several decays similar combinations of couplings with the same lepton flavour appear, e.g., $\pi^0,~\eta,~\eta^\prime,~\rho^0,~\omega$. For these cases the strongest bound on vector operators comes from the $\rho^0$ mode, while the $\eta^\prime$ channel provides a stronger limit on the scalar and tensor contributions.

\subsection{Leptonic dipole moments and rare decays of leptons}
\label{subsec:g-2}

\begin{figure}[th]
	\begin{center}
	\includegraphics[width=0.7\linewidth]{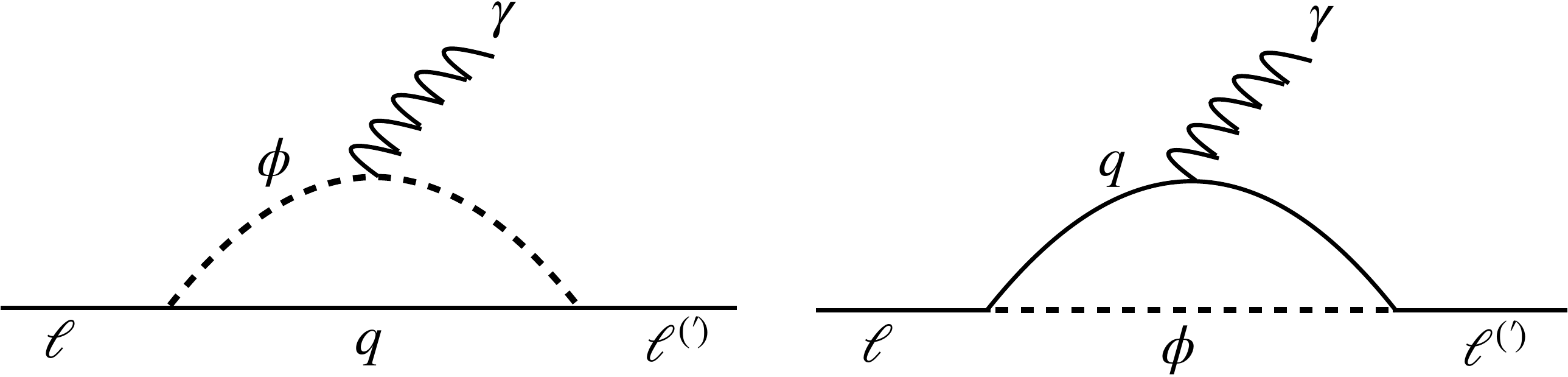}
	\caption{Scalar leptoquark ($\phi$) contributions to lepton dipole moments ($\ell' = \ell$) and $\ell\to\ell^\prime\gamma$.}
\label{dia:rare}
\end{center}
\end{figure}

The leptoquark coupling to a charged lepton and a quark can give rise to an anomalous magnetic or electric dipole moment of the corresponding charged lepton (when $\ell^\prime\equiv\ell$), or to the radiative lepton-flavour-violating decay  $\ell\to\ell^\prime\gamma$, via the one-loop diagrams shown in Fig.~\ref{dia:rare}.

\subsubsection{Anomalous magnetic moments}

The interaction term 
\begin{equation}
\label{eq:efflq}
\bar \ell_i ( \lambda_L^{ij} P_R + \lambda_R^{ij} P_L ) q_j \: 
\phi^* + {\rm h.c.} \, ,
\end{equation}
with $\phi$ being the scalar leptoquark and $\lambda^{ij}$ the corresponding Yukawa coupling, induces NP contributions to the anomalous magnetic moment $a_\ell \equiv\frac{1}{2}\left(g-2\right)_\ell$ given by~\cite{Choudhury:2001ad,Cheung:2001ip}
\begin{align}
\label{eq:g-2}
\Delta a_{\ell_i}\, =\, \frac{-3}{16 \pi^2} 
\frac{m_\ell^2}{M^2_{\rm LQ}}\, \sum_{j}\, &\bigg\{
( |\lambda_L^{ij}|^2 + |\lambda_R^{ij}|^2 )\, \left[ Q_{q_j} F_1(x_j) + Q_{\rm LQ} F_2(x_j) \right] \nn \\
&+ \frac{m_{q_j}}{m_\ell}\; \Re
(\lambda_L^{ij} \lambda_R^{ij*})\, \left[ Q_q F_3(x_j) + Q_{\rm LQ} F_4(x_j)\right]
\bigg\}\, ,
\end{align}
where the loop functions are defined as
\begin{eqnarray}
F_1(x_j) &=& 
\frac{1}{6 \,(1-x_j)^4} \, (2+3 \, x_j-6 \, x_j^2+x_j^3+6 \, x_j  \, \ln x_j) \;, \nn \\
F_2(x_j) &=& \frac{1}{6 \, (1-x_j)^4} \, (1-6 \, x_j+3 \, x_j^2+2 \, x_j^3-6 \, x_j^2 \, \ln x_j) \; ,
\nn \\
F_3(x_j) &=&
\frac{1}{(1-x_j)^3} \, (-3+4 \, x_j-x_j^2-2 \,  \ln x_j) \; ,
\nn \\
F_4(x_j) &=& \frac{1}{(1-x_j)^3} \, (1-x_j^2+2 \, x_j \,\ln x_j) \;.
\end{eqnarray}
In the above expression, $Q_q$ and $Q_{\rm LQ} $ are the EM charges of the quark and leptoquark flowing in the loop, respectively, $x_j=m_{q_j}^2/M^2_{\rm LQ}$, and we have neglected terms proportional 
to $m_\ell^2/M^2_{\rm LQ}$. Note that when working in the charge-conjugate quark basis one has to flip the sign of the mass and charge of the corresponding quark in the above expressions.

It is interesting to note that the current discrepancies between data and theoretical estimates for the muon and electron $g-2$ have opposite signs. The difference $\Delta a_\mu \equiv a_\mu^{\rm exp}-a_\mu^{\rm SM}$ is non-zero and positive with a significance of $3.7\sigma$~\cite{Bennett:2006fi,Davier:2017zfy,Jegerlehner:2017lbd,Keshavarzi:2018mgv,Blum:2018mom}, whereas the deviation is at the $2.4\sigma$ level for $\Delta a_e\equiv a_e^{\rm exp}-a_e^{\rm SM}$ and with the opposite sign~\cite{Hanneke:2008tm,Mohr:2015ccw,Aoyama:2017uqe,Parker:2018vye,Davoudiasl:2018fbb}.
The explicit values are quoted in the first column of Table~\ref{Table:g-2}.

It can be easily seen from Eq.~\eqref{eq:g-2} that leptoquarks having both left- and right-handed couplings to charged leptons can generate much larger contributions than those with only one type (either left or right) of interaction, due to the enhancement from the quark mass in the loop, especially the top quark. In that case, the second term in Eq.~\eqref{eq:g-2} dominates over the first term. Such scenario occurs for the $R_2$ and $S_1$ leptoquarks. After summing over the contributions from the second and third quark generations in the loop (the contribution from the first generation is negligible), the respective constraint equations for $R_2$ and $S_1$ can be written as
\begin{equation}
\label{eq:R2G-2}
\Re(y^\prime_{i3}z_{3i})+0.029 \,\Re(y^\prime_{i2}z_{2i})
\; =\; \left\{
\begin{array}{lcl}
\dsp (1.2\pm 0.5)\!\times\! 10^{-4}\;\left(\frac{M_{\rm LQ}}{\rm TeV}\right)^2 & & (i=1), 
\\[3ex]
\dsp \dsp (-1.8\pm 0.5)\!\times\! 10^{-3}\;\left(\frac{M_{\rm LQ}}{\rm TeV}\right)^2 & \hskip-20pt & (i=2),
\end{array}
\right.
\end{equation}
\begin{equation}
\label{eq:S1G-2}
\Re(\tilde{y}_{3i}z_{3i}^*)+0.042 \,\Re(\tilde{y}_{2i}z_{2i}^*)
\; =\;\left\{
\begin{array}{lcl}
\dsp (2.0\pm 0.8)\!\times\! 10^{-4}\;\left(\frac{M_{\rm LQ}}{\rm TeV}\right)^2 & & (i=1), 
\\[2ex]
\dsp (-3.0\pm 0.8)\!\times\! 10^{-3}\;\left(\frac{M_{\rm LQ}}{\rm TeV}\right)^2 &  & (i=2),
\end{array}
\right.
\end{equation}
where $i=1,2$ represent the electron and muon cases, respectively. As the loop functions depend on the leptoquark mass, it is not possible to 
completely factor out the dependence on $M_{\rm LQ}$.
The numerical coefficients written above have been obtained with $M_{\rm LQ}=1$~TeV.

These two equations depict the allowed $\pm 1\sigma$ regions that could explain the measured anomalous magnetic moments. In the first and second rows of Table~\ref{Table:g-2} we separately highlight the needed ranges of leptoquark couplings for the discrepancy to be fully ascribed to either the top or charm quark flowing in the loop, respectively. It can be noted from Eqs.~\eqref{eq:R2G-2} and \eqref{eq:S1G-2} that the difference in limits is not simply linear in quark masses, as the loop functions depend significantly on $m_{q_j}$. The explanation of the muon $g-2$ anomaly in explicit leptoquark models is subject to several other constraints; some detailed studies can be found in Refs.\cite{ColuccioLeskow:2016dox,Kowalska:2018ulj}.

\begin{table}[!h]
	\centering
	\renewcommand{\arraystretch}{1.2}
		\begin{tabular}{ c c  c}
			\hline \hline
			$\Delta a_\ell \equiv a_\ell^{\rm exp}-a_\ell^{\rm SM}$ & 
$R_2$ leptoquark & $S_1$ leptoquark
\\
			\hline \noalign{\vskip3pt}
			\multirow{2}{*}{$\Delta a_e = \left(-87\pm 36\right)\!\times\!10^{-14} $ }& $\Re(y^\prime_{13}z_{31})\!\in\! [7,17]\!\times\!10^{-5} $ & $\Re(\tilde{y}_{31}z_{31}^*)\!\in\! [12,28]\!\times\!10^{-5}$  \\[0.5ex]
			&  $\Re(y^\prime_{12}z_{21})\!\in\! [24,58]\!\times\!10^{-4} $ & $\Re(\tilde{y}_{21}z_{21}^*)\!\in\! [27,66]\!\times\!10^{-4}$  
\\[.2ex]\hline \noalign{\vskip 1pt}
			\multirow{2}{*}{$\Delta a_\mu = \left(2.74\pm 0.73\right)\!\times\!10^{-9}$} & $\Re(y^\prime_{23}z_{32})\!\in\! [-23,-13]\!\times\!10^{-4}$ & $\Re(\tilde{y}_{32}z_{32}^*)\!\in\![-37,-22]\!\times\!10^{-4}$\\[0.5ex] 
			&  $\Re(y^\prime_{22}z_{22})\!\in\! [-78,-45]\!\times\!10^{-3}$ & $\Re(\tilde{y}_{22}z_{22}^*)\!\in\![-88,-51]\!\times\!10^{-3}$\\[0.5ex] 
			\hline
			\hline 
	\end{tabular}
	\caption{$1\sigma$ ranges of $R_2$ and $S_1$ leptoquark couplings able to explain the electron and muon anomalous magnetic moments, for
$M_{\rm LQ}=1$~TeV. For larger leptoquark masses, the numbers increase approximately as $M_{\rm LQ}^2$.}
\label{Table:g-2}
\end{table}

In the absence of either the left- or right-handed coupling to charged leptons, the expression in Eq.~\eqref{eq:g-2} simplifies significantly and can be written, in the limit $x_j\to 0$, as
\begin{align}
\label{eq:g-2-2}
\Delta a_{\ell_i}\, &=\, \frac{-3}{16 \pi^2} 
\frac{m_\ell^2}{M^2_{\rm LQ}}\: \sum_{j}
|\lambda_{L/R}^{ij}|^2  \left[ Q_q F_1(x_j) + Q_{\rm LQ} F_2(x_j) \right], \nn  \\
&=\, \frac{-3}{96 \pi^2} 
\frac{m_\ell^2}{M^2_{\rm LQ}}\:\sum_{j} |\lambda_{L/R}^{ij}|^2 \left( 2\,Q_q  + Q_{\rm LQ}  \right).
\end{align}
Due to the $m_\ell$ suppression, the resulting ranges of couplings are irrelevant for a TeV-mass leptoquark, as they exceed the perturbativity limit.
Therefore, we do not show them in Table~\ref{Table:g-2} and simply conclude that the $\tilde{R_2},~\tilde{S_1}$ and $S_3$ leptoquarks cannot provide an explanation of the magnetic moment anomalies.\footnote{Ref.~\cite{Dorsner:2019itg} avoids the chiral suppression through scenarios which combine two different leptoquarks with fermionic couplings of opposite chirality.}

\subsubsection{Electric dipole moments}

Leptoquarks can also induce a lepton electric dipole moment (EDM) through the imaginary part of the Yukawa couplings in Eq.~\eqn{eq:efflq}. The effect is significant only when the leptoquark couples directly to both the left- and right-handed charged lepton, so that at one loop the top quark mass can induce the chirality flip. The relevant expression is given by~\cite{Cheung:2001ip}
\begin{align}
\label{eq:EDM}
|d_{\ell_i}|\; =\;  \frac{3\,e}{32 \pi^2}\: \sum_{j}
\frac{m_{q_j}}{M^2_{\rm LQ}}\;  \big|\Im
(\lambda_L^{ij} \lambda_R^{ij*}) \left[ Q_q F_3(x_j) + Q_{\rm LQ} F_4(x_j) \right]\big|\, .
\end{align}

The most stringent limit on the electron EDM, extracted from polar ThO molecules~\cite{Baron:2013eja}, is given in the second column of Table~\ref{Table:EDM}. This 90\% C.L. bound excludes several BSM models with time-reversal symmetry violating interactions and, as expected, the ensuing limits on the imaginary part of the product of leptoquark couplings (for $R_2$ and $S_1$) are also very restrictive. However the current bound on the muon EDM~\cite{Bennett:2008dy} gives a much weaker constraint on these NP couplings, with $\mathcal{O}(1)$ values still allowed.
The experimental EDM limits constrain combinations of couplings similar to the l.h.s of Eqs.~\eqref{eq:R2G-2} and \eqref{eq:S1G-2}, replacing the real parts by their imaginary parts. Instead of summing over contributions from all quarks, we separately show each contribution in Table~\ref{Table:EDM}, where the bounds on the top quark couplings are written in the first rows, for both electron and muon EDMs; however, for the charm quark  only the electron EDM provides a relevant bound, shown in the second row. A discussion on the constraints from EDMs of nucleons, atoms, and molecules on scalar leptoquark couplings can be found in Ref.~\cite{Dekens:2018bci}.

\begin{table}[!h]
	\centering
	\renewcommand{\arraystretch}{1.3}
		\begin{tabular}{ c c c  c}
			\hline \hline
			$|d_\ell|$& $|d_\ell^{\rm\, exp}|$ ($e$\,cm) & 
			$R_2$ leptoquark & $S_1$ leptoquark
 \\
			\hline \noalign{\vskip4pt}
			\multirow{2}{*}{$|d_e| $}   & \multirow{2}{*}{$<\, 8.7\times 10^{-29} $} & $|\Im(y^\prime_{13}z_{31})|<6.2 \times 10^{-10} $ & $|\Im(\tilde{y}_{31}z_{31}^*)|< 1.0\times 10^{-9}$  \\[0.5ex]
			& & $|\Im(y^\prime_{12}z_{21})|<2.2 \times 10^{-8} $ & $|\Im(\tilde{y}_{21}z_{21}^*)|< 2.4\times 10^{-8}$  
\\[.2ex]\hline \noalign{\vskip 1pt}
			$|d_\mu| $ & $<\, 1.9\times\!10^{-19}$ & $|\Im(y^\prime_{23}z_{32})|< 1.4$ & $|\Im(\tilde{y}_{32}z_{32}^*)|<2.1$\\[0.5ex] 
			\hline
			\hline 
	\end{tabular}
	\caption{Bounds on $R_2$ and $S_1$ leptoquark couplings from the electric dipole moments of leptons, at 90\% C.L. (95\% C.L.) for the electron (muon), for $M_{\rm LQ}=1$~TeV. For larger leptoquark masses, the numbers increase approximately as $M_{\rm LQ}^2$.}	
\label{Table:EDM}
\end{table}

\subsubsection[Radiative  $\ell \to \ell^\prime \gamma$ decays]{Radiative  $\boldsymbol{\ell \to \ell^\prime \gamma}$ decays}

The interaction term in Eq.~\eqref{eq:efflq} can also generate the rare lepton-flavour-violating decays $\ell \to \ell^\prime \gamma$. Apart from the two Feynman topologies shown in Fig.~\ref{dia:rare}, there exist two more diagrams where the photon is emitted from any of the external lepton lines. Including all four contributions, the decay width can be written as  \cite{Lavoura:2003xp,Benbrik:2008si}
\begin{eqnarray}
{\Gamma}(\ell_i \to \ell'_k \gamma) &=& \frac{\alpha}{4}\,
\frac{(m^2_\ell - m^2_{\ell'})^3}{ m^3_\ell }\;\sum_{j} \bigg(
|A_{R}^{ijk}|^2 + |A_{L}^{ijk}|^2 \bigg),
\end{eqnarray}
where
\begin{eqnarray}
\label{eq:AR}
A_{R}^{ijk} &=& \frac{3}{32 \pi^2}
\frac{1}{M^2_{\rm LQ}}\; \Bigg\{
\big(m_{\ell_i} \lambda_L^{ij} \lambda_L^{kj*}  + m_{\ell'_k}\lambda_R^{ij} \lambda_R^{kj*} \big)
\big[ Q_qF_{1}(x_j) + Q_{\rm LQ}F_{2}(x_j)\big] 
\nn \\ 
&&\hskip 1.95cm\mbox{}+ m_{q_j}\,
\big(\lambda_L^{ij} \lambda_R^{kj*}\big) \big[ Q_q F_{3}(x_j) +  Q_{\rm LQ}
F_4 (x_j)\big]\Bigg\}\, , 
\\
A_{L}^{ijk} &=& A_{R}^{ijk}( R \leftrightarrow L)\, . 
\end{eqnarray}
The terms proportional to $m_{\ell^{(\prime)}}$ arise from the topologies where the photon is emitted from the $\ell^{(\prime)}$ line. These contributions are suppressed compared to the enhancement due to heavy quarks flowing in the loop, as shown by the last term in Eq.~\eqref{eq:AR}. Similarly to the previous discussion of dipole moments, only the leptoquarks having both left- and right-handed couplings to charged leptons can generate such enhancement.

\begin{table}[!t]
	\centering
	\renewcommand{\arraystretch}{1.1}
			\resizebox{1.02\textwidth}{!}{\hspace*{-0.5cm}
	\begin{tabular}{c c c c}
		\hline \hline
LQ	&Bounds from $\mu \to e \gamma $  & Bounds from $\tau \to e \gamma $ & Bounds from $\tau \to \mu \gamma $  
\\
\hline \noalign{\vskip3pt}
		\multirow{2}{*}{$R_2 $}   & $|y^\prime_{23}z_{31}|^2,|y^\prime_{13}z_{32}|^2\!<\!1.2\times10^{-15}$ & $|y^\prime_{33}z_{31}|^2,|y^\prime_{13}z_{33}|^2\!<\!1.4\times10^{-7}$ &$|y^\prime_{33}z_{32}|^2,|y^\prime_{23}z_{33}|^2\!<\!1.9\times10^{-7}$ \\[0.5ex] 
		& $|y^\prime_{22}z_{21}|^2,|y^\prime_{12}z_{22}|^2\!<\!1.3\times10^{-12}$ & $|y^\prime_{32}z_{21}|^2,|y^\prime_{12}z_{23}|^2\!<\!1.6\times10^{-4}$ &$|y^\prime_{32}z_{22}|^2,|y^\prime_{22}z_{23}|^2\!<\!2.2\times10^{-4}$ 
\\[.2ex]\hline \noalign{\vskip 1pt} 
		\multirow{2}{*}{$S_1 $}   & $|\tilde{y}_{32}z_{31}^*|^2,|\tilde{y}_{31}z_{32}^*|^2\!<\!3.0\times10^{-15} $ & $|\tilde{y}_{33}z_{31}^*|^2, |\tilde{y}_{31}z_{33}^*|^2\!<\!3.8\times10^{-7}$ & $|\tilde{y}_{33}z_{32}^*|^2,|\tilde{y}_{32}z_{33}^*|^2\!<\!5.0\times10^{-7} $\\[0.5ex]
		& $|\tilde{y}_{22}z_{21}^*|^2,|\tilde{y}_{21}z_{22}^*|^2\!<\!1.7\times10^{-12} $ & $|\tilde{y}_{23}z_{21}^*|^2, |\tilde{y}_{21}z_{23}^*|^2\!<\!2.1\times10^{-4}$ & $|\tilde{y}_{23}z_{22}^*|^2,|\tilde{y}_{22}z_{23}^*|^2\!<\!2.8\times10^{-4} $
\\[.2ex]\hline \noalign{\vskip 1pt} 
		$\tilde{S_1} $   & $|y_{32}y_{31}^*|^2\!<\!5.4\times10^{-7} $ & $|y_{33}y_{31}^*|^2\!<\!2.3\times10^{-1}$ & $|y_{33}y_{32}^*|^2\!<\!3.1\times10^{-1} $
\\[.2ex]\hline \noalign{\vskip 1pt}		
		\multirow{2}{*}{$S_3 $}   & $|y_{32}y_{31}^*|^2\!<\!1.3\times10^{-7} $ & $|y_{33}y_{31}^*|^2\!<\!5.8\times10^{-2}$ & $|y_{33}y_{32}^*|^2\!<\!7.7\times10^{-2} $\\[0.5ex]
		& $|\tilde{y}_{32}\tilde{y}_{13}^\dagger|^2\!<\!3.4\times10^{-6} $ & $|\tilde{y}_{33}\tilde{y}_{13}^\dagger|^2\!<\!1.5$ & $|\tilde{y}_{33}\tilde{y}_{23}^\dagger|^2\!<\! 1.9 $\\[0.5ex]
		\hline
		\hline 
	\end{tabular}}
	\caption{Bounds on leptoquark couplings from $\ell \to \ell^\prime \gamma$, at 90\% C.L.\,. The limits are obtained for $M_{\rm LQ}=1$~TeV, and scale (approximately for $R_2 $ and $S_1$) as $M_{\rm LQ}^4$ for heavier leptoquark masses.}
\label{Table:rare}
\end{table}

The MEG experiment provides the most stringent upper limit on $\mu\to e\gamma$ \cite{TheMEG:2016wtm}, while for  $\tau \to  \ell\gamma$ the strongest bounds have been put by BaBar \cite{Aubert:2009ag}. The current 90\% C.L. limits are:
\begin{eqnarray}
\label{eq:datalgamma}
{\rm BR}(\mu\to e\gamma)< 4.2\times 10^{-13}, \quad
{\rm BR}(\tau\to e\gamma)< 3.3\times 10^{-8},\quad
{\rm BR}(\tau\to \mu\gamma)< 4.4\times 10^{-8}.
\end{eqnarray}
These experimental bounds imply the constraints on the appropriate combinations of leptoquark Yukawa parameters given in Table~\ref{Table:rare}.
As discussed above, due to the large top-loop contribution, we find severe limits for the $R_2$ and $S_1$ leptoquark couplings, as shown in their first row in the table. Whereas the second row for these two leptoquarks displays much weaker limits on the charm couplings (assuming that only the charm quark contributes in the loop). For $R_2$, the relevant contributions come from its component $R_2^{5/3}$, because the other charge component $R_2^{2/3}$ couples only to right-handed leptons as can be seen from Eq.~\eqref{eq:Lfull2}.  It is interesting to note in Table~\ref{Table:rare} that, contrary to the case of lepton dipole moments where reasonable bounds are absent for the leptoquarks with only left- or right-handed interactions, here,  in these rare decays we find significant upper bounds (especially in $\mu\to e \gamma$) for the $\tilde{S_1}$ and $S_3$ leptoquarks. For $S_3$, the first and second rows in the table correspond to  the limits arising from its $S_3^{4/3}$ and $S_3^{1/3}$ components, respectively. There are no useful bounds for $\tilde{R_2}$ because the corresponding combination of EM charges and loop functions, $Q_d F_{1}(x_j) + Q_{\tilde{R_2}} F_{2}(x_j)$, is almost vanishing for down-type quarks and a TeV-mass $\tilde{R_2}$.

\subsubsection[Rare $\ell\to\ell^\prime\ell^\prime\ell^{\prime\prime}$ decays]{Rare $\boldsymbol{\ell\to\ell^\prime\ell^\prime\ell^{\prime\prime}}$ decays}

\begin{figure}[h]
	\begin{center}
		\includegraphics[width=0.3\linewidth]{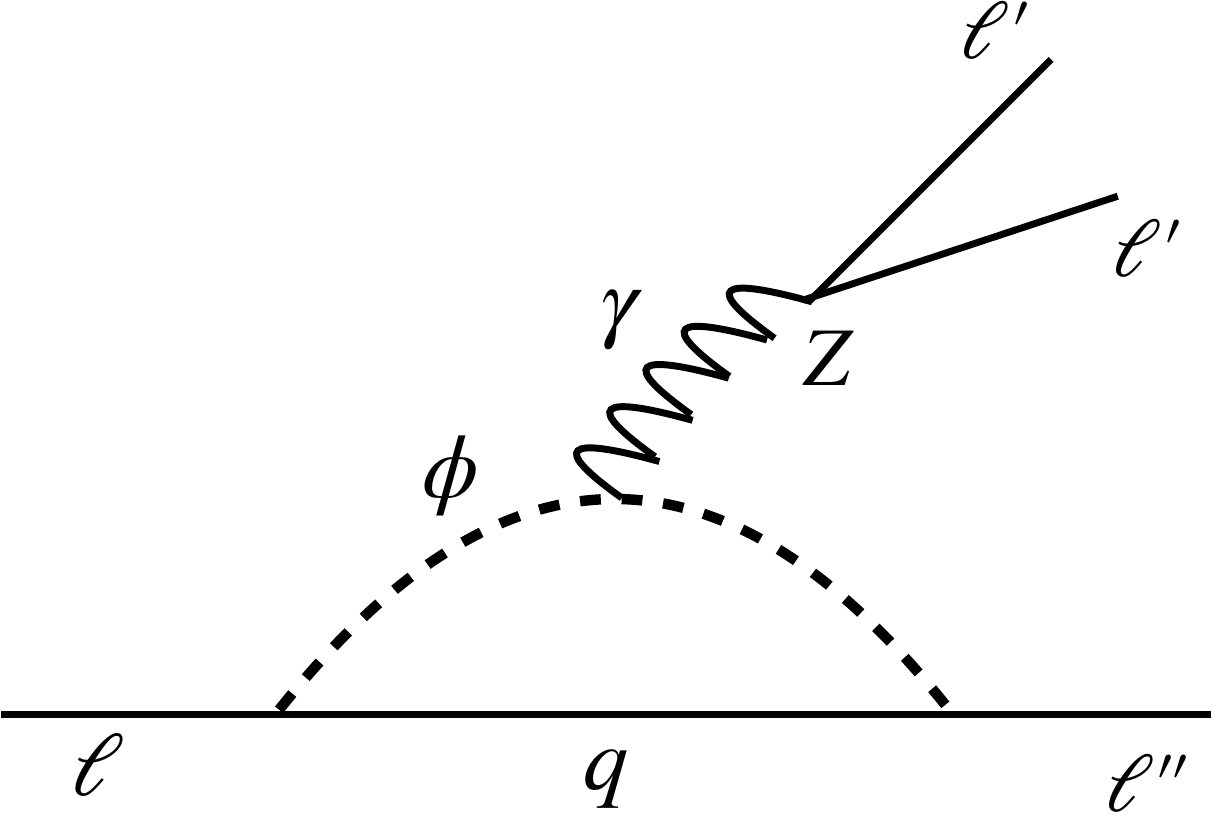}\hspace*{3cm}
		\includegraphics[width=0.27\linewidth]{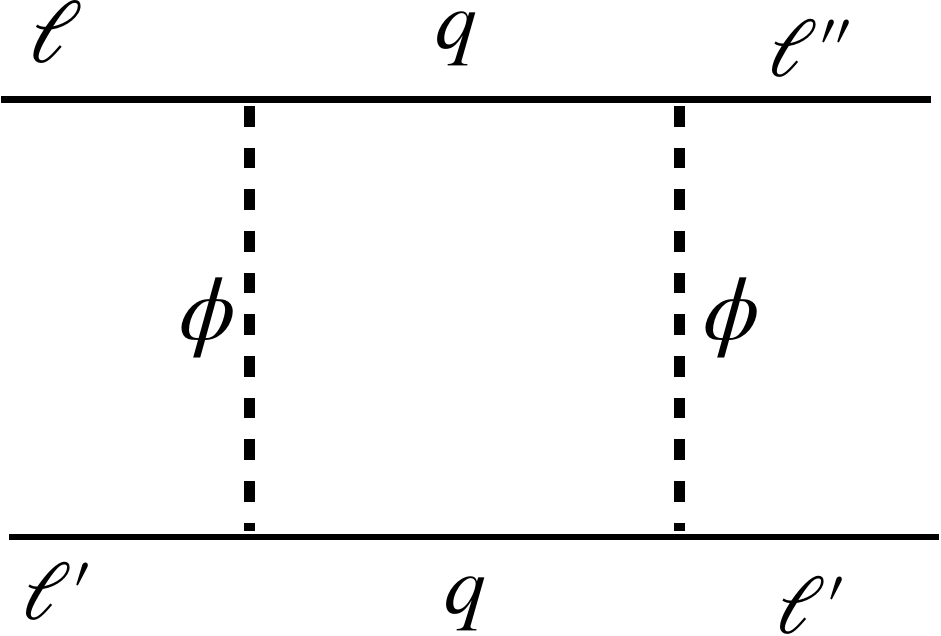}
		\caption{Penguin and box scalar leptoquark ($\phi$) contributions to the decays $\ell\to\ell^\prime\ell^\prime\ell^{\prime\prime}$. Diagrams with the leptoquark and quark lines interchanged are not shown.}\label{dia:3l}
	\end{center}
\end{figure}

The rare lepton-flavour-violating decays $\ell\to\ell^\prime\ell^\prime\ell^{\prime\prime}$ are also induced by the leptoquarks, at the one-loop level. These decays proceed via penguin diagrams with $Z$ and $\gamma$ exchanges, and via box diagrams with quarks and leptoquarks within the loop, as shown in the left and right panels of Fig.~\ref{dia:3l}, respectively.
The interaction term in Eq.~\eqref{eq:efflq} generates the 
following decay rate into final leptons with identical flavour~\cite{Arganda:2005ji,Benbrik:2010cf,deBoer:2015boa,Abada:2014kba}:
\begin{align}
\label{eq:3l}
{\rm  BR}(\ell_i^- \to (3\ell_n)^-) \;=\;\frac{\alpha_e^2m_{\ell_i}^5}{32\pi\Gamma_{\ell_i}}\;
&\bigg\{ |T_{1L}|^2+|T_{1R}|^2+\left(|T_{2L}|^2+|T_{2R}|^2\right)\left(\frac{16}{3}\ln\frac{m_{\ell_i}}{m_{\ell_n}}-\frac{22}{3}\right)\nonumber\\
&-4\,\mathrm{Re}[T_{1L}T_{2R}^*+T_{2L}T_{1R}^*]\nonumber\\
&+\frac16\left(|B_{1L}|^2+|B_{1R}|^2\right)+\frac13\left(|B_{2L}|^2+|B_{2R}|^2\right)\nonumber\\
&+\frac13\left(2\left(|Z_L g_{Ll}|^2+|Z_R g_{Rl}|^2\right)+|Z_L g_{Rl}|^2+|Z_R g_{Ll}|^2\right)\nonumber\\
&+\frac23\,\mathrm{Re}[T_{1L}B_{1L}^*+T_{1R}B_{1R}^*+T_{1L}B_{2L}^*+T_{1R}B_{2R}^*]\nonumber\\
&-\frac43\,\mathrm{Re}[T_{2R}B_{1L}^*+T_{2L} B_{1R}^* +T_{2L}B_{2R}^*+T_{2R} B_{2L}^*]\nonumber\\
&+\frac23\,\mathrm{Re}[B_{1L}Z_L^*g_{Ll}+B_{1R}Z_R^*g_{Rl}+B_{2L}Z_L^*g_{Rl}+B_{2R}Z_R^*g_{Ll}]\nonumber\\
&+\frac23\,\mathrm{Re}[2\, (T_{1L}Z_L^*g_{Ll}+T_{1R}Z_R^*g_{Rl})+T_{1L}Z_L^*g_{Rl}+T_{1R}Z_R^*g_{Ll}]\nonumber\\
&+\frac23\,\mathrm{Re}[-4\, (T_{2R}Z_L^*g_{Ll}+T_{2L}Z_R^*g_{Rl})-2(T_{2L}Z_R^*g_{Ll}+T_{2R}Z_L^*g_{Rl})]\bigg\}\,. 
\end{align}
This expression gets slightly modified when there are two different lepton flavours in the final state\cite{Abada:2014kba}~:
\begin{align}
\label{eq:llplp}
{\rm  BR}(\ell_i^- \to \ell_m^- \ell_n^- \ell_n^+) \;=\;\frac{\alpha_e^2m_{\ell_i}^5}{32\pi\Gamma_{\ell_i}}\;
&\bigg\{ \frac23 (|T_{1L}|^2+|T_{1R}|^2)+\left(|T_{2L}|^2+|T_{2R}|^2\right)\left(\frac{16}{3}\ln\frac{m_{\ell_i}}{m_{\ell_n}}-8\right)\nonumber\\
&-\frac83\,\mathrm{Re}[T_{1L}T_{2R}^*+T_{2L}T_{1R}^*]\nonumber\\
&+\frac{1}{12}\left(|B_{1L}|^2+|B_{1R}|^2\right)+\frac13\left(|B_{2L}|^2+|B_{2R}|^2\right)\nonumber\\
&+\frac13\left(|Z_L g_{Ll}|^2+|Z_R g_{Rl}|^2+|Z_L g_{Rl}|^2+|Z_R g_{Ll}|^2\right)\nonumber\\
&+\frac13\,\mathrm{Re}[T_{1L}B_{1L}^*+T_{1R}B_{1R}^*+2\,(T_{1L}B_{2L}^*+T_{1R}B_{2R}^*)]\nonumber\\
&-\frac23\,\mathrm{Re}[T_{2R}B_{1L}^*+T_{2L} B_{1R}^* +2\,(T_{2L}B_{2R}^*+T_{2R} B_{2L}^*)] \nonumber\\
&+\frac13\,\mathrm{Re}[B_{1L}Z_L^*g_{Ll}+B_{1R}Z_R^*g_{Rl}+2\,(B_{2L}Z_L^*g_{Rl}+B_{2R}Z_R^*g_{Ll})]\nonumber\\
&+\frac23\,\mathrm{Re}[ T_{1L}Z_L^*g_{Ll}+T_{1R}Z_R^*g_{Rl}+T_{1L}Z_L^*g_{Rl}+T_{1R}Z_R^*g_{Ll}]\nonumber\\
&-\frac43\,\mathrm{Re}[T_{2R}Z_L^*g_{Ll}+T_{2L}Z_R^*g_{Rl}+T_{2L}Z_R^*g_{Ll}+T_{2R}Z_L^*g_{Rl}]\bigg\}\,, 
\end{align}
The contributions from photon penguin diagrams are encoded in the $T_{1L,1R}$ and $T_{2L,2R}$ terms, whereas the $Z$-penguin effects are included in $Z_{L,R}$. The box-diagram decay amplitudes are denoted by $B_{1L,1R},\, B_{2L,2R}$. 
\tp{It} can be seen from the detailed expressions given in Eqs.~\eqref{eq:T1exp}--\eqref{eq:B2exp}, that the penguin contributions are enhanced by a factor $\ln(M_{\rm LQ}^2/m_{q_j}^2)$ and dominate over the box contributions, for leptoquark masses in the TeV range:
\begin{align}
\label{eq:T1exp}
T_{1L,1R}\; &=\; -\frac3{16\pi^2}\frac1{M^2_{\rm LQ}}\;\lambda_{L,R}^{ij} \lambda_{L,R}^{mj*} \left[\left(\frac49+\frac13\ln x_j\right)Q_q+\frac{1}{18}\, Q_{\rm LQ}\right]\,,\\
T_{2L,2R}\; &=\; -\frac3{16\pi^2}\frac1{M^2_{\rm LQ}}\,\bigg\{ \left[\frac16\,\lambda_{R,L}^{ij}\lambda_{R,L}^{mj*}+\frac{m_{q_j}}{m_{\ell_j}}\,\lambda_{R,L}^{ij}\lambda_{L,R}^{mj*}\left(\frac32+\ln x_j\right)\right] Q_q\nonumber\\
&\hskip 2.9cm\mbox{}
+\left(\frac1{12}\,\lambda_{R,L}^{ij}\lambda_{R,L}^{mj*}-\frac12\,\frac{m_{q_j}}{m_{\ell_i}}\,\lambda_{R,L}^{ij}\lambda_{L,R}^{mj*}\right) Q_{\rm LQ}\bigg\}\,,\\
Z_{L,R}\; &=\; -\frac3{16\pi^2}\frac1{M^2_{\rm LQ}}\;\lambda_{L,R}^{ij}\lambda_{L,R}^{mj*}\frac1{m_Z^2\sin^2\theta_w\cos^2\theta_w}
\nonumber\\
&\times\left[\frac34\, m_{\ell_i}^2\,
g_{Lq,Rq}-m_{q_j}^2\left(1+\ln x_j\right)g_{Rq,Lq}
- \frac34\, m_{\ell_i}^2\, g
\right]\, ,
\\
B_{1L,1R}\; &=\;\frac3{32\pi^2}\frac{-1}{M^2_{\rm LQ}}\;\lambda_{L,R}^{ij}\lambda_{L,R}^{mj*}\left|\lambda_{L,R}^{nk}\right|^2\,,\\
\label{eq:B2exp}
B_{2L,2R}\; &=\;\frac3{64\pi^2}\frac{-1}{M^2_{\rm LQ}}\;\lambda_{L,R}^{ij}\lambda_{L,R}^{mj*}\left|\lambda_{R,L}^{nk}\right|^2,
\end{align}
with
\begin{align}
g_{Ll,Rl}=T_3^{(l_L,l_R)}+ \sin^2\theta_w\,,\quad g_{Lq,Rq}=T_3^{(q_L,q_R)}-Q_q \sin^2\theta_w\,,\quad g=T_3^{\rm LQ}-Q_{\rm LQ}\sin^2\theta_w\,.
\end{align}
Here, $T_3^{\rm LQ}$, $T_3^{(l_L,l_R)}$ and $T_3^{(q_L,q_R)}$ denote the third components of the weak isospin of the leptoquark, the SM charged leptons and the quarks, respectively.

These rare decays have not been yet observed at experiments. The current 90\% C.L. upper bounds are~\cite{Tanabashi:2018oca}:
\begin{eqnarray}
\label{eq:data3l}
&{\rm BR}(\mu^- \to e^- e^-e^+)<1.0\times 10^{-12}\, , 
\nn \\
&{\rm BR}(\tau^- \to e^- e^-e^+)<2.7\times 10^{-8}\, ,
\qquad\qquad
{\rm BR}(\tau^- \to \mu^- \mu^-\mu^+)<2.1\times 10^{-8}\, , 
\nn \\
&{\rm BR}(\tau^- \to e^- \mu^- \mu^+)<2.7\times 10^{-8}\, ,
\qquad\qquad
{\rm BR}(\tau^- \to \mu^- e^-e^+)<1.8\times 10^{-8}\, ,
\nn \\
&{\rm BR}(\tau^- \to e^+ \mu^- \mu^-)<1.7\times 10^{-8}\, ,
\qquad\qquad
{\rm BR}(\tau^- \to \mu^+ e^-e^-)<1.5\times 10^{-8}\, .
\end{eqnarray}
For the first five modes both the penguin and box diagrams contribute, whereas the last two decays proceed only via box diagrams. We note that for the leptoquarks having both left- and right-handed couplings to quarks and leptons, i.e., $R_2$ and $S_1$, the resulting limits on the product of Yukawa couplings are two to three orders of magnitude weaker in the $\ell\to\ell^\prime\ell^\prime\ell^{\prime\prime}$ mode compared to the corresponding rare decay $\ell \to \ell^\prime \gamma$ (shown in Table~\ref{Table:rare}). Hence we do not quote such limits here. Instead we obtain constrained equations among various couplings of the form
\begin{align}
\label{eq:cons3l}
|\lambda_{L,R}^{ij} \lambda_{L,R}^{mj*}|^2 \left[ a_1^j+ a_2^j\, \sum_{k=1}^3\left|\lambda_{L,R}^{nk}\right|^2  +\left( \sum_{k=1}^3\left|\lambda_{L,R}^{nk}\right|^2\right)^{\!\!2}  \right] \le\, a_3\, .
\end{align}
\begin{table}[!t]
	\centering
	\renewcommand{\arraystretch}{1.3}
	\resizebox{1.05\textwidth}{!}{\hspace*{-0.5cm}
		\begin{tabular}{c| c c c c c }
			\hline \hline
			&$R_2$ & $S_1$ & $\tilde{R}_2$ & $\tilde{S}_1$ & $S_3$  \\
			& ($R_2^{5/3},R_2^{2/3}$)&&&& ($S_3^{4/3},S_3^{1/3}$)\\ \hline \hline
			\noalign{\vskip2pt}
			\multirow{2}{*}{$a_2^{1,2,3}$} & $-47.2,-24.6,8.0$ & $45.4,22.8,-9.8$ & $20.9,15.6,8.6$ &$-22.7,-17.4,-10.4$& $-22.7,-17.4,-10.4$ \\ 
			& $20.9,15.6,8.6$&&&& $45.4,22.8,-9.8$\\[.3ex]	\hline		
			\parbox[c]{1mm}{\multirow{7}{*}{\rotatebox[origin=c]{90}{$\mu^-\!\! \to 3\,e$}}} \multirow{2}{*}{$|\lambda_{L,R}^{ij} \lambda_{L,R}^{mj*}|^2$}
	 & $| y^\prime_{2j} y_{j1}^{\prime\,\dagger}|^2,|z_{j2}z_{j1}^*|^2$& $|\tilde{y}_{j2}\tilde{y}_{1j}^\dagger|^2,|z_{j2}z_{j1}^*|^2$ & $|y_{j2}y_{j1}^*|^2$ &$|y_{j2}y_{j1}^*|^2$ & $4|y_{j2}y_{j1}^*|^2$ \\
			&$| y_{2j} y_{j1}|^2$&&&& $|\tilde{y}_{j2}\tilde{y}_{1j}^\dagger|^2$\\[.5ex]
			\multirow{2}{*}{$\lambda_{L,R}^{nk}$} & $y^\prime_{1k},z_{k1}$ &$\tilde{y}_{k1},z_{k1}$&$y_{k1}$ & $y_{k1}$ & $\sqrt{2} y_{k1}$ \\
			& $y_{1k}$ &&& & $ \tilde{y}_{k1}$ \\[.3ex]
			\multirow{2}{*}{$a_1^{1,2,3}$} & $861.5,252.4,95.2$& $776.8,197.9,71.4$ & $164.1,91.0,27.7$ &  $204.7,124.4,51.8$& $204.7,124.4,51.8$\\
			& $164.1,91.0,27.7$ &&& & $776.8,197.9,71.4$\\[.3ex]
			$a_3$ &$2.9\times 10^{-3}$&$2.9\times 10^{-3}$&$2.9\times 10^{-3}$&$2.9\times 10^{-3}$&$2.9\times 10^{-3}$\\
			\hline	\noalign{\vskip2pt}
			\parbox[t]{1mm}{\multirow{9}{*}{\rotatebox[origin=c]{90}{$\tau^-\!\! \to 3\,e\; (\mu^- 2\,e)$}}}  \multirow{2}{*}{$|\lambda_{L,R}^{ij} \lambda_{L,R}^{mj*}|^2$}
& $| y^\prime_{3j} y_{j1}^{\prime\,\dagger}|^2,|z_{j3}z_{j1}^*|^2$& $|\tilde{y}_{j3}\tilde{y}_{1j}^\dagger|^2,|z_{j3}z_{j1}^*|^2$ & $|y_{j3}y_{j1}^*|^2$ &$|y_{j3}y_{j1}^*|^2$ & $4|y_{j3}y_{j1}^*|^2$ \\
			&$| y_{3j} y_{j1}|^2$&&&& $|\tilde{y}_{j3}\tilde{y}_{1j}^\dagger|^2$\\[.5ex]
			\multirow{2}{*}{$\lambda_{L,R}^{nk}$} & $y^\prime_{1k},z_{k1}$ &$\tilde{y}_{k1},z_{k1}$&$y_{k1}$ & $y_{k1}$ & $\sqrt{2} y_{k1}$ \\
			& $y_{1k}$ &&& & $ \tilde{y}_{k1}$ \\[.3ex]
			\multirow{4}{*}{$a_1^{1,2,3}$} & $884.1,275.0,117.8$ & $779.3,200.4,74.0$ & $164.1,91.0,27.7$ &  $214.7,134.4,61.8$ &  $214.7,134.4,61.8$\\
			&($1213.0,401.0,221.4$) &$(1042.9,271.1,124.0)$ &$(218.8,121.3,37.0)$  & ($301.5,194.5,97.8$) & ($301.5,194.5,97.8$) \\
			& $164.1,91.0,27.7$ &&& & $779.3,200.4,74.0$\\
			&$(218.8,121.3,37.0)$ & & & & $(1042.9,271.1,124.0)$ \\[.3ex]
			\multirow{2}{*}{$a_3$}  &$4.4\times 10^{2}$&$4.4\times 10^{2}$&$4.4\times 10^{2}$&$4.4\times 10^{2}$&$4.4\times 10^{2}$\\
			&$(5.8\!\times\! 10^{2})$&$(5.8\!\times\! 10^{2})$&$(5.8\!\times\! 10^{2})$&$(5.8\!\times\! 10^{2})$&$(5.8\!\times\! 10^{2})$ \\
			\hline  \noalign{\vskip2pt}
			\parbox[t]{3mm}{\multirow{9}{*}{\rotatebox[origin=c]{90}{$\tau^-\!\! \to  3\,\mu\; (e^- 2\,\mu)$}}}  \multirow{2}{*}{$|\lambda_{L,R}^{ij} \lambda_{L,R}^{mj*}|^2$}
& $| y^\prime_{3j} y_{j2}^{\prime\,\dagger}|^2,|z_{j3}z_{j2}^*|^2$& $|\tilde{y}_{j3}\tilde{y}_{2j}^\dagger|^2,|z_{j3}z_{j2}^*|^2$ & $|y_{j3}y_{j2}^*|^2$ &$|y_{j3}y_{j2}^*|^2$ & $4|y_{j3}y_{j2}^*|^2$ \\
			&$| y_{3j} y_{j2}|^2$&&&& $|\tilde{y}_{j3}\tilde{y}_{2j}^\dagger|^2$\\[.5ex]
			\multirow{2}{*}{$\lambda_{L,R}^{nk}$} & $y^\prime_{2k},z_{k2}$ &$\tilde{y}_{k2},z_{k2}$&$y_{k2}$ & $y_{k2}$ & $\sqrt{2} y_{k2}$ \\
			& $y_{2k}$ &&& & $ \tilde{y}_{k2}$ \\[.3ex]
			\multirow{4}{*}{$a_1^{1,2,3}$} & $841.3,232.3,75.0$ & $774.5,195.7,69.2$ & $164.1,91.0,27.7$ &  $195.7,115.4,42.8$ & $195.7,115.4,42.8$\\
			& ($1127.5,315.5,135.8$) & $(1033.4,271.1, 114.5)$ & $(218.8,121.3,37.0)$  &  ($263.5,156.5,59.7$) & ($263.5,156.5,59.7$)\\
			& $164.1,91.0,27.7$ &&& & $774.5,195.7,69.2$ \\
			&$(218.8,121.3,37.0)$ &&&& $(1033.4,271.1, 114.5)$  \\[.3ex]
			\multirow{2}{*}{$a_3$} &$3.4\!\times\! 10^{2}$ &$3.4\!\times\! 10^{2}$ &$3.4\!\times\! 10^{2}$& $3.4\!\times\! 10^{2}$&$3.4\!\times\! 10^{2}$\\
			&$(8.7\!\times\! 10^{2})$&$(8.7\!\times\! 10^{2})$&$(8.7\!\times\! 10^{2})$&$(8.7\!\times\! 10^{2})$&$(8.7\!\times\! 10^{2})$ \\
			\hline 
	\end{tabular}}
\caption{Coefficients of the constrained equation \eqref{eq:cons3l}, 
arising from $\ell\to\ell^\prime\ell^\prime\ell^{\prime\prime}$,	
for all five scalar leptoquarks and $M_{\rm LQ}=1\,$TeV.} 
\label{Table:3l}
\end{table}
Writing the constraints in this manner, we find that the coefficients $a_{1,2}^j$, where $j$ is the generation index of the quark going in the loop, depend on the corresponding quark mass whereas $a_3$ is independent of it. Here $k$ is the index of the other quark in the box diagram. The values of these coefficients are shown in Table~\ref{Table:3l}. It can be seen that $a_{2}^j$ are process independent and depend on the mass and quantum numbers of the leptoquark. For $R_2$ and $S_3$ we show the bounds separately for each component $R_2^{5/3},R_2^{2/3}$ and $S_3^{4/3},S_3^{1/3}$ in two consecutive rows. 
The numerical coefficients $a_{2}^j$ are one order of magnitude smaller than $a_1^j$, which indeed reflects that the box contributions are suppressed compared to the penguin terms. Also note that the logarithms of $x_j$ are large for light quarks and therefore the bounds are stronger for them, opposite to what was obtained in the $\ell\to \ell^\prime\gamma$ channel, as the loop functions are quite different. The
constraints extracted from $\mu\to3\,e$ are quite acceptable, e.g., $|\lambda_{L,R}^{ij} \lambda_{L,R}^{mj*}|^2\sim 10^{-5}$ in absence of $\lambda_{L,R}^{nk}$, whereas the $\tau$ modes fail to impose reasonable limits as almost $\mathcal{O}(1)$ values are permitted. This also holds true for the last two decay modes in Eq.~\eqref{eq:data3l}, which proceed only via box diagrams, where we find that the combination $\lambda_{L,R}^{ij}\lambda_{L,R}^{mj*}\left|\lambda_{R,L}^{nk}\right|^2$ is allowed up to $\sim2^4$ for leptoquark masses of $\cO(1\,\mathrm{TeV})$. Future improvements in data can be important to obtain limits on these coupling constants.

\subsection[$\mu-e$ conversion]{$\boldsymbol{\mu-e}$ conversion}
\label{subsec:mue}

Similarly to the lepton-flavour-violating decays discussed in the preceding section, muon conversion in nuclei is also another interesting process providing complementary sensitivity to NP. Currently the strongest
bound is found in the case of gold nuclei where the 90\% C.L. limit is set by the SINDRUM experiment as~\cite{Bertl:2006up}
\begin{align}
{\rm BR}^{\mathrm{Au}}_{\mu-e}\; =\; \frac{\Gamma (\mu^- {\rm Au} \to e^- {\rm Au})}{\Gamma_{\rm capture}} \;\le\; 7 \times 10^{-13}.
\end{align}
Here the muon capture rate for gold is $\Gamma_{\rm capture}= 8.6\times 10^{-18} $\,GeV~\cite{Suzuki:1987jf}.

The operators contributing to $\mu-e$ conversion within nuclei, arising from leptoquark interactions, are given in Eq.~\eqref{eq:LncLep} with $m=1$ and $n=2$.
There are additional contributions from dipole operators, namely $\bar{e}_{L,R}\, \sigma_{\mu\nu}\, \mu_{L,R} F^{\mu\nu}$, where $F^{\mu\nu}$ is the EM field strength tensor. However, the constraints on these dipole operators from $\mu-e$ conversion are one order of magnitude weaker than the bounds from $\mu\to e \gamma$ given in Table~\ref{Table:rare}, and hence we do not quote them here.

We use the results derived in Ref.~\cite{Kitano:2002mt}, where the conversion rate is given by
\begin{align}
\Gamma_{\rm conv}\, =\, 4 m_\mu^5\; \Big| \tilde{g}_{LS}^{(p)}\, S^{(p)}+\tilde{g}_{LS}^{(n)}\, S^{(n)} + \tilde{g}_{LV}^{(p)}\, V^{(p)}+\tilde{g}_{LV}^{(n)}\, V^{(n)} \Big|^2 + {( L \to R)},
\end{align}
with the coupling constant $\tilde{g}$'s defined as
\begin{align}
\tilde{g}_{LS,RS}^{(p)}\;=\; &\sum_{q} G_S^{(q,p)} \frac{1}{2}[g_{S,q}^{R,L}]^{ii,12}\, ,\\
\tilde{g}_{LS,RS}^{(n)}\;=\; &\sum_{q} G_S^{(q,n)} \frac{1}{2}[g_{S,q}^{R,L}]^{ii,12}\, ,  \end{align}
\begin{align}
\tilde{g}_{LV}^{(p)}\;=\; &([g^{LL}_{V,u}]^{11,12}+[g^{RL}_{V,u}]^{11,12} )+\frac{1}{2}\, ([g^{LL}_{V,d}]^{11,12}+[g^{RL}_{V,d}]^{11,12})\, , \\
\tilde{g}_{RV}^{(p)}\;=\; &([g^{RR}_{V,u}]^{11,12}+[g^{LR}_{V,u}]^{11,12} )+\frac{1}{2}\, ([g^{RR}_{V,d}]^{11,12}+[g^{LR}_{V,d}]^{11,12})\, , \\
\tilde{g}_{LV}^{(n)}\;=\; &\frac{1}{2}\, ([g^{LL}_{V,u}]^{11,12}+[g^{RL}_{V,u}]^{11,12} )+([g^{LL}_{V,d}]^{11,12}+[g^{RL}_{V,d}]^{11,12})\, , \\
\tilde{g}_{RV}^{(n)}\;=\; &\frac{1}{2}\, ([g^{RR}_{V,u}]^{11,12}+[g^{LR}_{V,u}]^{11,12} )+([g^{RR}_{V,d}]^{11,12}+[g^{LR}_{V,d}]^{11,12})\, .
\end{align}
The overlap-integral values are
$S^{(p)}=0.0523$, $S^{(n)}=0.0610$, $V^{(p)}=0.0859$ and $V^{(n)}=0.108$, and the coefficients for scalar operators are evaluated as $G_S^{u,p}=G_S^{d,n}=5.1,~G_S^{d,p}=G_S^{u,n}=4.3$ and $G_S^{s,p}=G_S^{s,n}=2.5$~\cite{Kosmas:2001mv}.

Using all these inputs we depict in Table~\ref{Table:mue} the extracted bounds on the product of leptoquark Yukawa elements. It can be seen from Eq.~\eqref{eq:S1op}--\eqref{eq:S33op}, that only the $R_2,S_1$ and $S_3$ leptoquarks couple to the $u$ quark and charged leptons. These couplings, for the vector and axial-vector operators, are shown in the first two rows of Table~\ref{Table:mue}, where quite strong bounds are visible. The third row displays those combinations of couplings for vector and axial-vector operators dealing with $d$ quarks and charged leptons where bounds are stronger. The last row shows the contributions from scalar operators, which arise only for the $R_2$ and $S_1$ leptoquark couplings to the $u$ quark and charged leptons, and provide the strongest limit.

\begin{table}[!h]
	\centering
	\renewcommand{\arraystretch}{1.07}
	\resizebox{1.05\textwidth}{!}{\hspace*{-0.5cm}
		\begin{tabular}{c c c c c}
			\hline \hline
			  &  & & & Bound \\
			$R_2$ & $S_1$ & $\tilde{R}_2,~\tilde{S}_1$ & $S_3$ &$\times  \left(M_{\rm LQ}/{\rm TeV}\right)^4 $ \\[0.5ex] \hline
			\noalign{\vskip2pt}
			 $|0.14\,y_{12}^{\prime\,\dagger} y_{11}^\prime\!+\! 0.15\,y_{11} y_{21}^*|^2$ &  &  & $|0.14\,\tilde{y}_{12}\tilde{y}^\dagger_{11} \!+\! 0.30\, y_{12}y_{11}^* |^2$& $4.7\times 10^{-11}$ \\[1.ex]
			 $|z_{12} z_{11}^*|^2$ & $|\tilde{y}_{12}\tilde{y}^\dagger _{11}|^2,\,|z_{12} z_{11}^*|^2$ & & & $2.4\times 10^{-11}$ \\[1.5ex]			 
			  & & $|y_{12} y_{11}^*|^2$& &$2.1\times 10^{-11}$ \\[1.5ex]
			 $|y_{12}^{\prime\,\dagger} z_{11}^*|^2,\,|y_{11}^{\prime\,\dagger} z_{12}|^2$ & $|\tilde{y}_{12} z_{11}^*|^2,\,|\tilde{y}_{11}^{\dagger} z_{12}|^2$& & &$6.7\times 10^{-12}$ \\[.2ex]
			\hline
			\hline 
		\end{tabular}}
	\caption{Bounds on leptoquark couplings from muon conversion to electron in gold nuclei.}  \label{Table:mue}
\end{table}

\section{Bounds from kaons}
\label{sec:Kaon}

Some of the rare (semi)leptonic decays of $K$ mesons are mediated by FCNCs and thus are suppressed in the SM.  Although most of these processes are dominated by long-distance contributions and significant efforts are devoted to sharpen the SM predictions \cite{Cirigliano:2011ny}, these modes are also important in constraining BSM interactions.\footnote{Correlations between leptoquark contributions to $\varepsilon'/\varepsilon$  and rare kaon decays have been investigated in Ref.~\cite{Bobeth:2017ecx}.}
This is achievable due to the
strong suppression of the SM decay amplitude $\mathcal{A}_{\rm SM}$, as well as the improvements in experimental sensitivity.
In the next three subsections we discuss the effect of NP operators, arising from scalar leptoquarks, in $K\to \ell_i^-\ell_j^+,~ \pi \ell_i^-\ell^+_j$ and $ \pi \nu \bar{\nu}$.
The total amplitude for these decays can be written as
\begin{equation}
\mathcal{A}=\mathcal{A}_{\rm SM} + \mathcal{A}_{\rm LQ}\, .
\end{equation}
Owing to the conservation of lepton flavour, $\mathcal{A}_{\rm SM}=0$ when $\ell_i\not=\ell_j$ up to tiny contributions proportional to neutrino masses.

\subsection{Rare leptonic decays of kaons}
\label{sec:Kaonlep}

The rare decays $K_{L,S}^0\to \ell^+\ell^-$ are forbidden at tree level in the SM. However, leptoquarks can contribute to these modes at lowest order, which imposes strong constraints on their coupling constants. The neutral-current operators with down-type quarks given in Eq.~\eqref{eq:LncLep} lead to such decays. 
The SM amplitude $\mathcal{A}_{\rm SM}$ is dominated by the long-distance contribution arising from a two-photon intermediate state:
$K_{S,L}^0\to \gamma^* \gamma^* \to \ell^+ \ell^-$ \cite{Ecker:1991ru}.
The estimated $K_S^0$ branching ratios are  \cite{Cirigliano:2011ny}:
\begin{align}
{\rm BR}^{\mathrm{LD}}(K_S^0\to e^+ e^-) = 2.1\times 10^{-14} 
\qquad\quad \mathrm{and} \qquad\quad
{\rm BR}^{\mathrm{LD}}(K_S^0\to \mu^+ \mu^-) = 5.1\times 10^{-12} \, .
\end{align}
In the SM, there exists a small $CP$-violating short-distance contribution
to $K_S^0\to \mu^+ \mu^-$ that is one order of magnitude smaller: 
${\rm BR}^{\mathrm{SD}}(K_S^0\to \mu^+ \mu^-) \simeq 1.7\times 10^{-13}$ \cite{Buchalla:1995vs,Isidori:2003ts}. Owing to its helicity suppression
($\mathcal{A}_{\rm SM}^{\mathrm{SD}}\propto m_\ell$),
the analogous short-distance contribution to the electron mode is completely negligible. The current experimental upper bounds on the electron \cite{Ambrosino:2008zi} and muon \cite{Aaij:2017tia,LHCb:2019aoh} modes, shown in Table~\ref{Table:Kaon}, are larger than their predicted SM values by five and two orders of  magnitude, respectively. Hence, to constrain the leptoquark couplings, we can neglect the SM contributions and assume that the leptoquark amplitudes saturate the experimental limits.

It can be seen from Eqs.~\eqref{eq:S1op}--\eqref{eq:S33op} that 
$S_1$ cannot contribute at tree level to these transitions, while for each of the other four scalar leptoquark types the contribution to $\mathcal{A}_{\rm LQ}$ is generated by a single (axial)vector operator with Wilson coefficient  $g_{V,d}^{XY}$, where $X,Y\in \{L,R\}$. The corresponding decay rate of the $P^0_{ki}\equiv q_k \bar{q}_i$ meson is given by 
\begin{align}\label{eq:Pll-decay}
\Gamma_{\rm LQ}(P^0_{ki} \to \ell_n^+ \ell_m^-)\, =\, \frac{f_P^2 \,\big|[g_{V,d}^{XY}]^{ik,mn} \big|^2}{64 \pi m_P^3}\, \lambda^{1/2}(m_P^2,m_{\ell_m}^2,m_{\ell_n}^2)  \left[  m_P^2( m_{\ell_m}^2\!+\!  m_{\ell_n}^2)- (m_{\ell_m}^2\! -  m_{\ell_n}^2)^2 \right]. 
\end{align}
The relevant coupling combinations are obviously
$\left\{[g_{V,d}^{XY}]^{21,mn} (1+\bar\epsilon_K)\mp [g_{V,d}^{XY}]^{12,mn} (1-\bar\epsilon_K)\right\}/\sqrt{2}$
for the $K_{S,L}^0$ decays, although we will neglect the small $CP$-violating admixture $\epsilon_K$.
We can see from Table~\ref{Table:Kaon} that for the $K_S^0\to e^+e^-$ mode, $\mathcal{O}(1)$ couplings are allowed, due to the explicit lepton-mass suppression in \eqn{eq:Pll-decay},
while for $K_S^0\to \mu^+\mu^-$ we get strong limits on the respective couplings.

\begin{table}[tb]
	\centering
	\renewcommand{\arraystretch}{1.1}
	\begin{tabular}{ c c c c c}
		\hline \hline
		&  &  &    &  Bound \\
		Modes & ${\rm BR_{exp}}$ & $R_2$ & $\tilde{R}_2,~\tilde{S}_1,~4\!\times\! S_3$ &  $\times \left(M_{\rm LQ}/{\rm TeV}\right)^2$ \\
		\hline \noalign{\vskip4pt}
		$K_S^0 \to e^+e^- $   & $<9.0\times10^{-9} $ & $|\Im(y_{11}y_{12}^*)|$ & $|\Im(y_{11}^*y_{21})|$ &  $2.0$ \\[1.5ex]
		$K_S^0 \to \mu^+\mu^- $   & $< 2.1\times 10^{-10} $ & $|\Im[(y_{21}y_{22}^*)|$ & $|\Im(y_{12}^*y_{22})|$ &  $1.6\times 10^{-3}$ \\[1.5ex]
		$K_L^0 \to e^+e^- $ & $9^{+6}_{-4}\times10^{-12}$ & $|\Re(y_{11}y_{12}^*)|$ & $|\Re(y_{11}^*y_{21})|$   & $2.0\times 10^{-3}$ \\[1.5ex]
		$K_L^0 \to \mu^+\mu^-$ &$(6.84\pm0.11)\times10^{-9}$ & $|\Re(y_{21}y_{22}^*)|$ & $|\Re(y_{12}^*y_{22})|$ & $4.7\times 10^{-5}$ \\[1.5ex]
		$K_L^0 \to e^\pm \mu^\mp$ & $<4.7\times 10^{-12}$ & $|y_{21}y_{12}^*+ y_{11}^*y_{22}|$ & $|y_{21}y_{12}^*+ y_{11}y_{22}^*|$  &  $1.9\times 10^{-5}$ \\[1.5ex]
		\hline
		\hline 
	\end{tabular}
	\caption{Limits on leptoquark couplings from leptonic kaon decays. The experimental upper bounds are at 90\% C.L.\,.}  \label{Table:Kaon}
\end{table}

The situation is a bit different for the observed decay modes $K_L^0\to \ell^+ \ell^-$. The absorptive long-distance contribution \cite{GomezDumm:1998gw} nearly saturates the precisely measured $BR(K_L^0\to \mu^+ \mu^-)$ \cite{Ambrose:2000gj}, leaving little room for the dispersive component which would include both the leptoquark and short-distance SM contributions. The long-distance prediction for the electron mode \cite{Cirigliano:2011ny} is also in agreement with the experimental value \cite{Ambrose:1998cc}, the tiniest branching ratio ever measured, although the uncertainties are much larger in this case. In order to impose bounds on the leptoquark couplings, we allow them to saturate the $1\sigma$ experimental uncertainties, which gives the limits quoted in Table~\ref{Table:Kaon}.

Since only a single $g_{V,d}^{XY}$ coupling can generate the $K^0\to\ell^+\ell^-$ transition, the tree-level leptoquark exchange gives rise to an helicity-suppressed pseudoscalar leptonic amplitude $\bar u_\ell\gamma_5 v_\ell$. Therefore, the final lepton pair is produced in a s-wave configuration (${}^1S_0$) that is odd under $CP$, implying that the $K^0_S$  leptoquark amplitude violates $CP$, while the $K^0_L$ transition preserves this symmetry. Both decays are then complementary since they constrain the imaginary and real parts, respectively, of the relevant combination of leptoquark couplings.

For the lepton-flavour-violating decay  $K_L^0\to \mu^\pm e^\mp$, an stringent upper bound is obtained for the corresponding leptoquark couplings, as no SM contribution exists for this mode.

\subsection{Rare semileptonic decays of kaons}
\label{sec:Kaonsemi}
In the SM, the FCNC semileptonic decay $K^+\to \pi^+\ell^+\ell^-$ is completely dominated by the $CP$-conserving amplitude arising from virtual photon exchange, $K^+\to \pi^+\gamma^*$, which is a vector contribution \cite{Ecker:1987qi}. There exist short-distance $Z$-penguin and $W$-box contributions, involving also axial-vector lepton couplings, but they are negligible in the total decay rate (three orders of magnitude smaller for the muon mode).
Adopting the usual parameterization for the $K(k)\to \pi(p)$ hadronic matrix element, 
\begin{align}
\langle \pi^-|\bar{d} \gamma^\mu s|K^- \rangle\, &=\, -\langle \pi^+|\bar{s} \gamma^\mu d|K^+ \rangle\, =\, (k+p)^\mu\, f_+^{K\pi}(q^2) + (k-p)^\mu\, f_-^{K\pi}(q^2) \, , 
\nn \\
f_-^{K\pi}(q^2)\, &=\, \frac{m_K^2-m_\pi^2}{q^2}\, \left[f_0^{K\pi}(q^2)-f_+^{K\pi}(q^2)\right]  ,
\end{align}
where $q^2 \equiv(k-p)^2$,
and including the leptoquark contribution proportional to $g_{V,d}^{XY}$,
the differential decay distribution for $K^+(k) \to \pi^+(p)\ell^+(q_1)\ell^-(q_2)$  is given by
\begin{align}
\label{eq:Kpill}
\frac{d\Gamma}{dz} \big(K^\pm \to \pi^\pm \ell^+_m\ell^-_m\big)\, &=\; \frac{G_F^2\alpha^2m_K^5}{12\pi(4\pi)^4}\, \sqrt{\bar{\lambda}}\; \sqrt{1-4\frac{r_\ell^2}{z}} 
\nn \\ & \times\; \Bigg\{\bar \lambda\,\bigg(1+2\frac{r_\ell^2}{z} \bigg) \bigg[|V_+(z)|^2 + \frac{2 \pi}{G_F\, \alpha}\,\Re\left(V_+^*(z)\, [g_{V,d}^{XY}]^{21,mm}\right)\, f_+^{K\pi}(z)\bigg] 
\nn \\ 
& + \frac{2 \pi^2}{G_F^2\, \alpha^2}\,\big|[g_{V,d}^{XY}]^{21,mm}\big|^2 \bigg[\bar{\lambda}\,\bigg(1-\frac{r_\ell^2}{z} \bigg)\, [f_+^{K\pi}(z)]^2 +\frac{3r_\ell^2}{z}\big(1-r_\pi^2)^2\, [f_0^{K\pi}(z)]^2 \bigg]  \Bigg\}\,, 
\end{align}
where we have used the dimensionless variable $z\equiv q^2/m_K^2$ and $\bar{\lambda}\equiv\lambda(1,z,r_\pi^2)$ with $r_i=m_i/m_K$.

The SM vector contribution is usually defined as~\cite{Cirigliano:2011ny}
\begin{align}
A_\text{V}^{K^+\to \pi^+\ell^+\ell^-} &=\, -\frac{G_F\alpha}{4\pi}\, V_+(z)\;\bar{u}_\ell(q_2)(\slashed{k}+\slashed{p}) v_\ell(q_1)\,. 
\end{align}
where the vector form factor $V_+(z)$ vanishes at $\mathcal{O}(p^2)$ in chiral perturbation theory ($\chi$PT) and can be parametrized as \cite{Ecker:1987qi,DAmbrosio:1998gur}
\begin{equation}
V_+(z) = a_+ + b_+z + V_+^{\pi\pi}(z)\,,
\label{V_dec}
\end{equation}
which is valid to $\cO(p^6)$. 
The unitary loop correction $V_+^{\pi\pi}(z)$ that contains the $\pi\pi$ re-scattering contributions can be obtained from Refs.~\cite{Cirigliano:2011ny,Bijnens:2002vr}. The parameters $a_+$ and $b_+$ encode local contributions from $\chi$PT low-energy constants, which at present can only be estimated in a model-dependent way~\cite{Cirigliano:2011ny}.

Integrating over the allowed phase space, $4r_\ell^2 \le z\le (1-r_\pi)^2$,
and using PDG~\cite{Tanabashi:2018oca} inputs for all parameters, we
obtain the following numerical expressions for the branching fractions: 
\begin{align}
\label{eq:eeNeq}
{\rm BR}(K^+ \to \pi^+e^+e^-) =  10^{-8} \times\, &\big[  0.1 + 58.9\, a_+^2 + 1.7\,b_+^2 + 15.9 \,a_+ b_+ - 3.2\, a_+ - 0.8\, b_+ \nn \\
&+\,  5.8\times 10^{4} \,|\tilde{g}_1|^2  \nn \\
&+  (-58.4 + 2.2\times 10^{3}\, a_+ + 2.9\times 10^{2}\, b_+ )\,\Re\,\tilde{g}_1 + 4.5\, \Im\,\tilde{g}_1\big]\, ,  
\\[5pt]
\label{eq:mumuNeq}
{\rm BR}(K^+ \to \pi^+\mu^+\mu^-) = 10^{-9} \times\, &\big[  1.1 + 117.6\, a_+^2 + 10.3\,b_+^2 + 67.7\,a_+ b_+ - 19.1\, a_+ - 6.3\, b_+ \nn \\
&+\,  2.7\times 10^{5} \,|\tilde{g}_2|^2  \nn \\
&+  (-3.5\!\times\! 10^{2}\! + 4.3\!\times\! 10^{3}\, a_+\! +\! 1.2\!\times\! 10^{3}\, b_+ )\,\Re\,\tilde{g}_2 + 41.1\, \Im\,\tilde{g}_2\big]\, . 
\end{align}
Here $\tilde{g}_m\equiv 2\,[g_{V,d}^{XY}]^{21,mm}\times  (\mathrm{1\,TeV})^2$ 
for the electron ($m=1$) and muon ($m=2$) modes, respectively. The explicit form of $\tilde{g}_m$ in terms of Yukawa elements 
can easily be read from Eqs.~\eqref{eq:R21op}--\eqref{eq:S33op},
for the four leptoquark types giving tree-level contributions:
\begin{eqnarray}
\tilde{S_1}:y_{1m}y_{2m}^*\, ,
\qquad 
R_2:y_{m1}y_{m2}^*\,,
\qquad 
\tilde{R_2}: y_{1m}^*y_{2m}\,,
\qquad 
S_3: 2\,y_{1m}y_{2m}^*\, ,
\end{eqnarray}
times a factor $(1~\mathrm{TeV}/ M_{\rm LQ})^2$.

The experimental branching fractions for these two modes~\cite{Tanabashi:2018oca} 
are given in Table~\ref{Table:Kaon2}.
In absence of any NP contributions to Eqs.~\eqref{eq:eeNeq} and \eqref{eq:mumuNeq}, the parameters $a_+$ and $b_+$ have been extracted 
from a fit to the measured $z$ distribution by NA48/2 \cite{Batley:2009aa,Batley:2011zz}:
\begin{eqnarray}
a_+^{ee}=-0.578\pm 0.016,~ b_+^{ee}=-0.779\pm 0.066,~
a_+^{\mu\mu}=-0.575\pm 0.039,~ b_+^{\mu\mu}=-0.813\pm 0.145\,. 
\end{eqnarray}
Leptoquarks would introduce two more real parameters in the fit.
However, due to the limited statistics available in these modes, the full fit (including the NP couplings) may not be worth to impose bounds on these couplings. While $\mathcal{O}(1)$ values are expected, in the SM, for $a_+$ and $b_+$, it can be seen from Eqs.~\eqref{eq:eeNeq} and \eqref{eq:mumuNeq} that for
$\tilde g_m\sim\cO(1)$ the tree-level leptoquark contribution would be three orders of magnitude larger than the contributions arising from these two parameters. 
Hence we take a conservative approach and determine the bounds on the NP couplings quoted in Table~\ref{Table:Kaon2}, by neglecting the SM effects, i.e., assuming that the leptoquark contribution alone saturates the measured branching fractions.

The lepton-flavour-violating modes $K^+\to\pi^+\mu^\pm e^\mp$ do not receive any SM contribution. The differential decay widths induced by the corresponding leptoquark-mediated amplitudes are given by
\begin{align}
\label{eq:Kpimue}
\frac{d\Gamma}{d z}\big(K^+\to\pi^+\mu^\pm e^\mp\big)\, &=\, 
\frac{m_K^5 }{48(4\pi)^3}\;\big|[g_{V,d}^{XY}]^{21,mn} \big|^2\; 
\sqrt{\bar{\lambda}}\; \bigg(1-\frac{r_\mu^2}{z}\bigg)^2
\nn \\
&\times 
\bigg\{\bar\lambda\,\bigg(2+\frac{r_\mu^2}{z}\bigg)\, [f_+^{K\pi}(z)]^2 + \frac{3r_\mu^2}{z}\,\big(1-r_\pi^2\big)^2\, [f_0^{K\pi}(z)]^2\bigg\}\,.
\end{align}
with $m,n\in\{1,2\}$ and $m\ne n$.
We use the stringent upper bound on BR$(K^+\to\pi^+\mu^+ e^-)$, from the BNL E865 experiment \cite{Sher:2005sp}, to constrain the leptoquark couplings.
After integrating over the allowed phase space, $r_\mu^2 \le z\le (1-r_\pi)^2$, 
this gives the 90\% C.L. limit quoted in Table~\ref{Table:Kaon2}.

\begin{table}[thb]
	\centering
	\renewcommand{\arraystretch}{1.1}
	\begin{tabular}{ c c c c c}
		\hline \hline
		 &  &  &    &  Bound (or range) \\
		Modes& ${\rm BR_{exp}}$ & $R_2$ & $\tilde{R}_2,~\tilde{S}_1,~4\!\times\! S_3$ &  $\times \left(M_{\rm LQ}/{\rm TeV}\right)^2$ \\
		\hline \noalign{\vskip4pt}
		$K^+ \to \pi^+e^+e^- $   & $(3.00\pm 0.09)\times 10^{-7} $ & $
		|y_{11}y_{12}^*|$ & $|y_{11}^*y_{21}|$ &  $2.3\times 10^{-2}$ \\[1.5ex]
		$K^+ \to \pi^+\mu^+\mu^- $   & $(9.4\pm 0.6)\times10^{-8} $ &  $|y_{21}y_{22}^*|$ & $|y_{12}^*y_{22}|$ &  $1.9\times 10^{-2}$
\\[1.5ex]
		$K^+ \to \pi^+\mu^+ e^- $   & $<1.3\times10^{-11} $ &  $|y_{21}y_{12}^*|$, $|y_{11}y_{22}^*|$  &  $|y_{21}y_{12}^*|$, $|y_{11}^*y_{22}|$ &   $1.9\times 10^{-4}$ \\[1.5ex]\hline \noalign{\vskip3pt}
		$K_S^0 \to \pi^0 e^+e^- $ & $(5.8^{+2.9}_{-2.4})\times10^{-9} $
& $|\Re(y_{11}y_{12}^*)|$ & $|\Re(y_{11}^*y_{21})|$  &  $3.1\times 10^{-2}$ 
\\[1.5ex]
		$K_S^0 \to \pi^0 \mu^+\mu^-$ &$(2.9^{+1.5}_{-1.2})\times10^{-9}$ &  $|\Re(y_{21}y_{22}^*)|$ &  $|\Re(y_{12}^*y_{22})|$ &  $3.3\times 10^{-2}$ 
\\[1.5ex]\hline \noalign{\vskip3pt}
		\multirow{2}{*}{$K_L^0 \to \pi^0 e^+e^- $} & \multirow{2}{*}{$<2.8\times10^{-10}$} &  $\Im(y_{11}y_{12}^*)$&  $\Im(y_{11}^*y_{21})$ (for $\tilde{S}_1$)    & $[-4.1,\,2.6]\!\times\! 10^{-4}$ \\[1.ex]
&&& $\Im(y_{11}^*y_{21})$& $[-3.6,\,2.9]\!\times\! 10^{-4}$ \\[1.5ex]
\multirow{2}{*}{$K_L^0 \to \pi^0 \mu^+\mu^-$} &\multirow{2}{*}{$<3.8\times10^{-10}$} &  $\Im(y_{21}y_{22}^*)$ &  $\Im(y_{12}^*y_{22})$ (for $\tilde{S}_1$)  &  $[-6.5,\,5.1]\!\times\! 10^{-4}$ \\[1.ex]
&&& $\Im(y_{12}^*y_{22})$ &  $[-5.8,\,5.7]\!\times\! 10^{-4}$ \\[1.5ex]
		$K_L^0 \to\pi^0  e^\pm \mu^\mp$ & $<7.6\times 10^{-11}$ &  $|(y_{21}y_{12}^* - y_{11}^*y_{22})|$ & $|(y_{21}y_{12}^* - y_{11}y_{22}^*)|$  &   $2.9\times 10^{-4}$ \\[1.5ex]
		\hline
		\hline 
	\end{tabular}
	\caption{90\% C.L. bounds on leptoquark couplings from rare semileptonic kaon decays.}  \label{Table:Kaon2}
\end{table}

The decays $K_S^0 \to \pi^0 \ell^+\ell^-$ are very similar to $K^+\to \pi^+\ell^+\ell^-$. Their differential decay distribution can be directly obtained from Eq.~\eqn{eq:Kpill}, replacing the vector form factor $V_+(z)$ by
$V_S(z) = a_S + b_S z + V_S^{\pi\pi}(z)$, and $[g_{V,d}^{XY}]^{21,mm}$ by the appropriate combination of leptoquark couplings 
$\tilde g^+_m\equiv ([g_{V,d}^{XY}]^{21,mm} + [g_{V,d}^{XY}]^{12,mm})/\sqrt{2}$.
The branching fractions take then the numerical form:
\begin{align}
\label{eq:eeNeq2}
{\rm BR}(K_S \to \pi^0e^+e^-) =  10^{-10}\times\, & \big[  0.02 + 46.90\, a_S^2 + 1.45\,b_S^2 + 13.03 \,a_S b_S - 0.79\, a_S - 0.25\, b_S \nn \\
+&\,  8.50\times 10^{4}\, |\tilde{g}_1^+|^2
\nn \\
+&  (-20.59 + 2.45\times\! 10^{3}\, a_S + 3.39\!\times\! 10^{2} \, b_S )\,\Re\,\tilde{g}_1 \!+\! 3.90 \,\Im\,\tilde{g}_1^+  \big] ,  
\end{align}
\begin{align}
\label{eq:mumuNeq2}
{\rm BR}(K_S \to \pi^0\mu^+\mu^-) = 10^{-11} \times\, &\big[  0.15 + 101.85\, a_S^2 + 9.13\,b_S^2 + 59.18\,a_S b_S - 5.98\, a_S - 2.04\, b_S \nn \\
+&\,  3.91\times 10^{5}\, |\tilde{g}_2^+|^2
\nn \\
+&  (-1.56\!\times\! 10^{2}\! + 5.32\!\times\! 10^{3}\, a_S\! +\! 1.54\!\times\! 10^{3}\, b_S )\,\Re\,\tilde{g}_2 + 35.55 \,\Im\,\tilde{g}_2^+  \big] . 
\end{align}

As the $K_S\to 2\pi$ modes saturate more than 99\% of the total $K_S$ decay width and only branching fraction measurements are available for $K_S^0 \to \pi^0 \ell^+\ell^-$, it is not possible to extract the two form factor
parameters from data. Assuming the vector-meson-dominance relation 
$b_S/a_S=1/r_V^2\approx0.4$ \cite{DAmbrosio:1998gur}, the NA48 data \cite{Batley:2003mu,Batley:2004wg} imply
\begin{eqnarray}
|a_S^{ee}|=1.06^{+0.26}_{-0.21}\, ,
\qquad\qquad 
|a_S^{\mu\mu}|=1.54^{+0.40}_{-0.32}\, .
\end{eqnarray}
Similarly to the $K^+\to \pi^+ \ell^+\ell^-$ mode, we neglect the SM contributions and obtain the 90\% C.L. limits shown in Table~\ref{Table:Kaon2}.

Next we move to the decay $K_L^0\to \pi^0 \ell^+ \ell^-$, which is an interesting mode as it receives contributions from three different mechanisms within the SM \cite{Ecker:1987hd}: an indirect $CP$-violating amplitude due to the $K^0-\bar{K^0}$ oscillation, a direct $CP$-violating transition induced by short-distance physics and a $CP$-conserving contribution from $K_L^0\to \pi^0 \gamma\gamma\to \pi^0\ell^+ \ell^-$. 
The relevant $g_{V,d}^{XY}$ leptoquark coupling generates a $K^0\to\pi^0\ell^+\ell^-$ amplitude with vector ($1^{--}$), axial-vector ($1^{++}$) and pseudoscalar ($0^{-+}$) leptonic structures, giving rise to a $CP$-even 
$\pi^0\ell^+\ell^-$ final state. Therefore, the $K_L^0\to\pi^0\ell^+\ell^-$ leptoquark amplitude violates $CP$.
Combining the two $CP$-violating SM amplitudes with the leptoquark contribution, the differential distribution can be written as
\begin{align}
\label{eq:KLCPV}
\frac{d\Gamma }{d z}\big(K_L^0\to \pi^0 \ell^+ \ell^-\big)_{CPV}&=\frac{G_F^2\alpha^2 m_K^5}{12\pi(4\pi)^4}\; \sqrt{\bar{\lambda}}\; \sqrt{1-4\frac{r_\ell^2}{z}}\nn \\
&\times\Bigg\{\bar{\lambda}\,\bigg(1+\frac{2r_\ell^2}{z} \bigg) \bigg[|V_0(z)|^2+ \frac{\sqrt{2} \pi}{G_F\alpha}\,\Re[V_0^*(z)\,\tilde g_m^-]
\, f_+^{K\pi}(z) +|A_0(z)|^2\bigg]\nn \\
&+6r_\ell^2\,\big(2+2r_\pi^2-z\big)\,|A_0(z)|^2+ \frac{3}{2}\, r_\ell^2 z\, |P_0(z)|^2    
- 6r_\ell^2\,\big(1-r_\pi^2\big)\,\Re\big[A_0(z)^*P_0(z)\big] \nn \\
& + \frac{ \pi^2}{G_F^2 \alpha^2}\,\big|\tilde g_m^-
\big|^2\, \bigg[\bar{\lambda}\,\big(1-\frac{r_\ell^2}{z} \big)\, [f_+^{K\pi}(z)]^2 +\frac{3r_\ell^2}{z}\,\big(1-r_\pi^2)^2\, [f_0^{K\pi}(z)]^2 \bigg] 
\nn \\
& + s_Y\,\frac{\sqrt{2} \pi}{G_F\alpha}\,\Re[A_0^*(z)\,\tilde g_m^-]
\, \bigg[\bar{\lambda}\,\big(1-\frac{4 r_\ell^2}{z} \big)\, f_+^{K\pi}(z) +\frac{6r_\ell^2}{z}\,\big(1-r_\pi^2)^2 f_0^{K\pi}(z) \bigg] 
\nn\\
& - s_Y\,\frac{\sqrt{2} \pi}{G_F\alpha}\,\Re[P_0^*(z)\,\tilde g_m^-]\,
3 r_\ell^2 (1-r_\pi^2)\, f_0^{K\pi}(z)
\Bigg\}\,,
\end{align}
where we have defined $\tilde g_m^- \equiv\dsp
([g_{V,d}^{XY}]^{21,mm} - [g_{V,d}^{XY}]^{12,mm})/\sqrt{2}$.
The factor $s_Y$ accounts for the different sign of the axial leptonic coupling in right-handed ($s_R=+1$) and left-handed ($s_L=-1$) currents.
The SM vector, axial-vector and pseudoscalar amplitudes for $K_L^0(k)\to \pi^0(p) \ell^+(q_1) \ell^-(q_2)$ are defined as
\begin{align}
A_\text{V}^{K_L\to \pi^0\ell^+\ell^-} &= -\frac{G_F\alpha}{4\pi}\; V_0(z)\;\bar{u}_\ell(q_2)(\slashed{k}+\slashed{p}) v_\ell(q_1)\,,\notag\\
A_\text{A}^{K_L\to \pi^0\ell^+\ell^-} &= -\frac{G_F\alpha}{4\pi}\; A_0(z)\;\bar{u}_\ell(q_2)(\slashed{k}+\slashed{p})\gamma_5 v_\ell(q_1)\,,\notag\\
A_\text{P}^{K_L\to \pi^0\ell^+\ell^-} &= + \frac{G_F\alpha}{4\pi}\; P_0(z)\, m_\ell\;\bar{u}_\ell(q_2)\gamma_5 v_\ell(q_1)\,.
\end{align}

The indirect $CP$-violating contribution is related to the $K_S^0 \to \pi^0 \ell^+ \ell^-$ amplitude, which is fully dominated by its vector component:
\begin{equation}
V_0^\text{indirect}(z)\, =\, \epsilon_K\; \left[ a_S+b_S z + V_S^{\pi\pi}(z)\right]\,\approx\, \epsilon_K\, a_S\,\Big(1+\frac{z}{r_V^2}\Big)\,,
\end{equation}
where $\epsilon_K\sim e^{i\pi/4}\, |\epsilon_K|$ parametrizes $K^0$--$\bar K^0$ mixing with $|\epsilon_K|=(2.228\pm 0.011)\times 10^{-3}$. 
In the SM, the direct $CP$-violating contributions 
are given by
\begin{align}
V_0^\text{direct}(z)&=i\,\frac{2\pi\sqrt{2}}{\alpha}\, y_{7V}\, f_+^{K\pi}(z)\,\Im\lambda_t\,,\notag\\
A_0^\text{direct}(z)&=i\,\frac{2\pi\sqrt{2}}{\alpha}\, y_{7A}\,
f_+^{K\pi}(z)\,\Im\lambda_t\,,\notag\\
P_0^\text{direct}(z)&=-i\,\frac{4\pi\sqrt{2}}{\alpha}\, y_{7A}\,
f_-^{K\pi}(z)\,\Im\lambda_t\,,
\end{align}
where $\lambda_t=V_{ts}^*V_{td}$. We use the estimates of the $K_{3\ell}$ form factors $f_\pm^{K\pi}(z)$ from Ref.~\cite{Antonelli:2008jg} and the Wilson coefficients $y_{7V,7A}$ from Ref.~\cite{Buchalla:2003sj}.


For the electron mode the $CP$-conserving contribution is estimated to be one order of magnitude smaller than the $CP$-violating one, while for the muon channel both of them are similar in magnitude with a slightly larger $CP$-violating amplitude~\cite{Cirigliano:2011ny}. 
However the experimental 90\% C.L. upper bounds for these modes are still $\mathcal{O}(10^{-10})$ \cite{AlaviHarati:2003mr,AlaviHarati:2000hs}, which is one order of magnitude above their SM  estimates. Hence, in order to constrain the leptoquark couplings, we ignore the $CP$-conserving SM contributions; 
this is a conservative attitude, since they do not interfere with the $CP$-violating amplitudes in the decay rate. The SM $CP$-violating contributions and their interference with the leptoquark couplings are fully taken into account in our numerical analysis, which gives the allowed range for couplings depicted in Table~\ref{Table:Kaon2}.

\begin{figure}[!h]
	\begin{center}
		\includegraphics[width=0.42\linewidth]{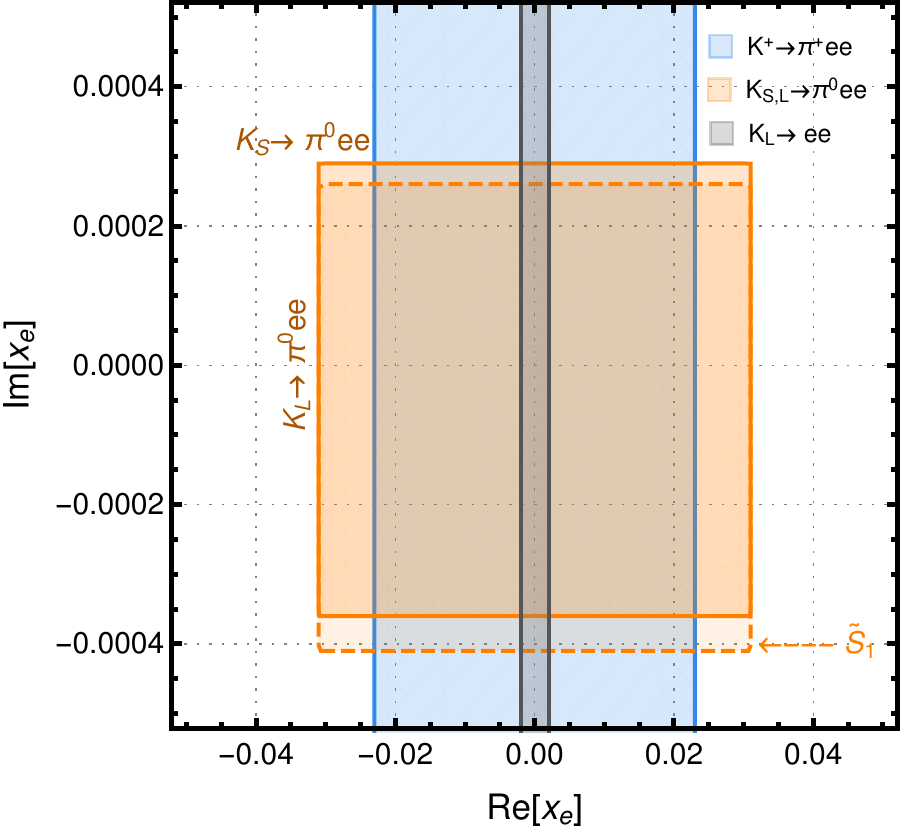} \hskip 25pt
		\includegraphics[width=0.42\linewidth]{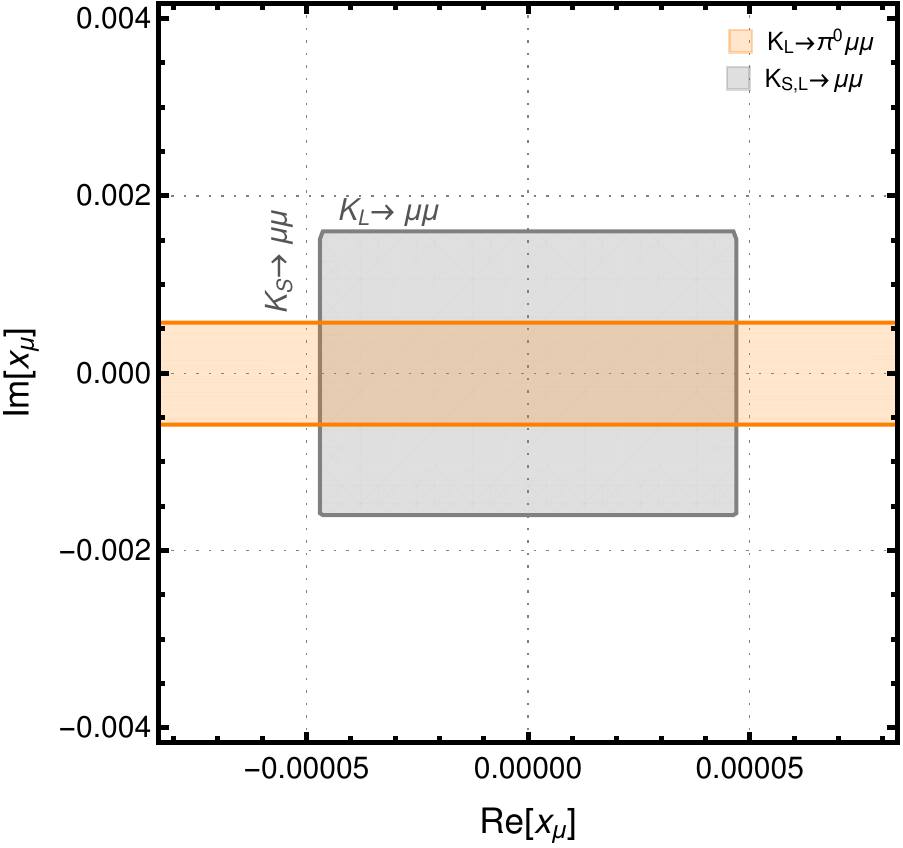}
		\caption{Allowed regions in the plane $(\mathrm{Re}[x_\ell], \mathrm{Im}[x_\ell])$,
arising from leptonic and rare semileptonic kaon decays, for the electron (left panel) and muon (right panel) channels. }\label{fig:Kaon}
	\end{center}
\end{figure}

It can be seen from Tables~\ref{Table:Kaon} and \ref{Table:Kaon2} that similar couplings are involved in these leptonic and rare semileptonic decay modes of kaons. The $K_S^0$ and $K_L^0$ decays are complementary, providing separate access to both the real and imaginary parts of the NP couplings, while the decays of the charged kaon restrict their absolute value. We highlight the situation in Fig.~\ref{fig:Kaon}, for both electron (left panel) and muon (right panel) modes, separately, where 
\begin{equation}
x_e = \left(\frac{1~\mathrm{TeV}}{M_{\mathrm{LQ}}}\right)^2\times\left\{
\begin{array}{c}
y_{11}^{\phantom{*}} y_{12}^* \\[2pt] y_{11}^{\phantom{*}} y_{21}^*
\end{array} \right.\, ,
\qquad\qquad
x_\mu = \left(\frac{1~\mathrm{TeV}}{M_{\mathrm{LQ}}}\right)^2\times\left\{
\begin{array}{c}
y_{21}^{\phantom{*}} y_{22}^* \\[2pt] y_{12}^{\phantom{*}} y_{22}^*
\end{array} \right.\, .
\end{equation}
The first line in the brackets corresponds to the leptoquark $R_2$, while the second line refers to $\tilde R_2,\, \tilde S_1,\, 4\times S_3$, as indicated in Tables~\ref{Table:Kaon} and \ref{Table:Kaon2}. 
The decay modes of the long-lived neutral kaon, $K_L^0\to\ell^+\ell^-,\pi^0\ell^+\ell^-$, put obviously more stringent constraints that their $K_S^0\to\pi^0\ell^+\ell^-,\ell^+\ell^-$ counterparts.
Other kaon decay modes, such as $K\to \ell\nu,\,\pi \ell\nu,\, \pi \pi \ell\nu$, with much larger SM contributions, cannot provide limits on the leptoquark couplings competitive with the ones extracted from rare decays.

Similarly to the $K^+$ case, the lepton-flavour-violating decays $K_L^0\to\pi^0\mu^\pm e^\mp$ have much simpler expressions, being mediated only by the leptoquark contribution. Their differential branching fractions are given by Eq.~\eqn{eq:Kpimue}, replacing $[g_{V,d}^{XY}]^{21,mm}$ by the appropriate combination of leptoquark couplings 
$([g_{V,d}^{XY}]^{21,mn}-[g_{V,d}^{XY}]^{12,mn})/\sqrt{2}$.
We notice in the last row of Table~\ref{Table:Kaon2} that a very stringent constraint arises from the current experimental upper limit on these modes \cite{Abouzaid:2007aa}.
The lepton-flavour-violating decays are absent in the SM and have not yet been seen in experiments. Hence the corresponding combinations of NP couplings have to be strictly suppressed to obey the experimental upper bounds.

\subsection{\boldmath $K\to\pi\nu\bar\nu$}
\label{sec:Kaonnunu}

Let us now consider the short-distance dominated decays $K\to \pi \nu \bar{\nu}$, which are thus expected to serve as very clean modes to look for BSM effects. 
These decay modes receive contributions from similar leptoquark couplings, but they involve three generations of neutrinos and the PMNS rotation has to be included suitably. 
In the presence of the leptoquark-induced operators with left-handed neutrinos and down-type quarks in Eq.~\eqn{eq:LncNeu}, the branching fractions for $K^+$ and $K_L^0$ can be written as
\begin{align}
\label{eq:Kppinunu}
{\rm BR}\left(  K^{+}\rightarrow\pi^{+}\nu\bar{\nu}
\right)\,   &  =\, \frac{\kappa_{\nu}^{+}}{3\, |V_{us}|^{10}}\,\left(  1+\Delta_{EM}\right)\,
\left\{ \sum_{\ell=1}^3\,
\left| y_{\nu} - \frac{ \pi\, \sin^2{\theta_{W}}}{\sqrt{2}\, G_F\, \alpha} \,
[N_{V_X}^{d}]^{21,\ell\ell}  \right|^2
\right.\nonumber\\ &\hskip 4.2cm\left.\mbox{}
+ \frac{ \pi^2\, \sin^4{\theta_{W}}}{2\, G_F^2\, \alpha^2} 
\,\sum_{m\not= n}\, \left| [N_{V_X}^{d}]^{21,mn} \right|^2\right\}\, ,
%
\\[2ex]
\label{eq:KLpinunu}
{\rm BR}\left(K_{L}^0\rightarrow\pi^{0}\nu\bar{\nu}\right)\,  &=\,
\frac{\kappa_{\nu}^{L}}{3\, |V_{us}|^{10}}\, \left( 1-\delta_\epsilon\right)\,
\left\{ \sum_{\ell=1}^3\,
\left| \Im\left( y_{\nu} - \frac{ \pi\, \sin^2{\theta_{W}}}{\sqrt{2}\, G_F\, \alpha} \,  [N_{V_X}^{d}]^{21,\ell\ell}\right)
\right|^2
\right.\nonumber\\ &\hskip 3.6cm\left.\mbox{}
+ \frac{ \pi^2\, \sin^4{\theta_{W}}}{8\, G_F^2\, \alpha^2} 
\,\sum_{m\not= n}\, \left| [N_{V_X}^{d}]^{21,mn} - [N_{V_X}^{d}]^{12,mn} \right|^2\right\}\, ,
%
\end{align}
where $X\in \{L,R\}$ and we have summed over all possible undetected neutrinos in the final state. 
In the last expression we have made use of the hermiticity of the Lagrangian $\cL_{\mathrm{eff}}^{\mathrm{nc},\nu}$ in Eq.~(\ref{eq:LncNeu}), which implies $([N_{V_X}^{q}]^{ik,mn})^* = [N_{V_X}^{q}]^{ki,nm}$.
The overall factors
\begin{equation}
\kappa_{\nu}^{+,\,L}   =\tau_{+,L}\,\frac{G_{F}^{2}\,\alpha  ^{2}\,m_{K^{+,0}}^{5}}{256\,\pi^{5}\sin^{4}\theta_{W}}\,  \left|  V_{us}\right|^{8}
\left|  V_{us}\times f_{+}^{K^i\pi^i}\left(  0\right)  \right|^{2}\, \mathcal{I}_{\nu}^{+,0} 
\end{equation}
are extracted from $K_{\ell 3}$ data. They encode the hadronic matrix element information,
with $\mathcal{I}_{\nu}^{+,0}$ the phase-space integral over the normalized vector form factor:
\begin{align}
\mathcal{I}_{\nu}^{i}  &  =\int_{0}^{\left(  1-r_{\pi}\right)  ^{2}}%
dz\; \bar{\lambda}^{3/2}\,\left|  \frac{f_{+}^{K^{i}\pi^{i}}\left(  z\right)  }%
{f_{+}^{K^{i}\pi^{i}}\left(  0\right)  }\right|^{2}\, ,
\qquad\qquad
i\in \{+,0\}.
\end{align}
The Wilson coefficient $y_\nu$ is given by
\begin{equation}
y_{\nu}=\left(  \mathrm{Re}\lambda_{t}+i\,\mathrm{Im}\lambda_{t}\right)
X_{t}+\left|  V_{us}\right|  ^{4}\mathrm{Re}\lambda_{c}\,P_{u,c}\;,
\label{NUCoeff}%
\end{equation}
with $\lambda_{c}=V_{cs}^{\ast}V_{cd}$, $X_{t}=1.464\pm0.041$ and $P_{u,c}=0.41\pm0.04$ \cite{Buras:2005gr,Buras:2006gb}. The electromagnetic correction
takes the value $\Delta_{EM} = -0.003$~\cite{Mescia:2007kn} and $\delta_\epsilon\simeq0.03$ accounts for the small
$K^0-\bar{K^0}$ mixing contribution~\cite{Cirigliano:2011ny,Buchalla:1996fp}.

Using PDG values~\cite{Tanabashi:2018oca} for all other inputs, we quote the constraint on the leptoquark couplings arising from the decay $K^{+}\rightarrow\pi^{+}\nu\bar{\nu}$  as
\begin{align}
\label{eq:KpnunuCons}
\sum_{\ell=1}^3 \bigg|\left(-23.0 + i\, 6.6\right)\times 10^{-5}- \tilde{N}_\ell \bigg|^2 + \sum_{m\not= n} |\tilde{N}_{mn}|^2 \pm  1.5   
\times 10^{-8} \le 3.6\times 10^{-7}\, ,
\end{align}
where 
the last term in the left-hand side accounts for the uncertainty on the SM prediction, while the right-hand side reflects the 
(90\% C.L.) upper bound $\mathrm{BR}(K^+ \to\pi^+  \nu \bar{\nu}) < 1.85\times 10^{-10}$, recently reported by the NA62 collaboration \cite{NA62_kaon2019}.
The parameters
$\tilde{N}_\ell\equiv 2\,[N_{V_X}^{d}]^{21,\ell\ell} \times  (\mathrm{1\,TeV})^2$ and $\tilde{N}_{mn}\equiv 2\,[N_{V_X}^{d}]^{21,mn} \times  (\mathrm{1\,TeV})^2$ 
contain the leptoquark couplings for identical and different neutrino flavours in the final state,
respectively. The explicit expressions of these couplings for the three relevant types of leptoquarks are quoted in Table~\ref{Table:KaonNu}, where we separately show, in the first and second rows, the allowed ranges for their real and imaginary parts for each neutrino generation $\ell$, and, in the third row, the bounds on the moduli of products of couplings with different neutrino flavours.
Due to the interference with the SM contribution, we find allowed ranges for the leptoquark couplings when the two final neutrinos have the same flavour, whereas an upper bound is obtained for different flavours. 
Since the SM predictions are very accurately known, the resulting bounds on the NP couplings are quite stringent, as can be seen from Table~\ref{Table:KaonNu}.

\begin{table}[t]
	\centering
	\renewcommand{\arraystretch}{1.05}
	\begin{tabular}{ c c c c}
		\hline \hline
		&    &     &  Range (or bound) \\
		Modes& ${\rm BR_{exp}}$ & $S_1, ~ S_3, ~ \tilde{R}_2$ &  $\times \left(M_{\rm LQ}/{\rm TeV}\right)^2$ \\
		\hline \noalign{\vskip4pt}

		\multirow{3}{*}{$K^+ \to\pi^+  \nu \bar{\nu}$ }&  \multirow{3}{*}{$< 1.85\times 10^{-10}$} 
		& $s_{\mbox{\tiny LQ}}\,\Re(\hat{y}_{1\ell}\, {\hat{y}}_{2\ell}^*)$
	& $[-3.7,\,8.3]\times 10^{-4}$ \\[0.5ex]
		& & $\Im(\hat{y}_{1\ell}\, {\hat{y}}_{2\ell}^*)$
	& $[-5.3,\,6.7]\times 10^{-4}$ \\[0.5ex]
		&& $\big[\!\sum \limits_{m\not=n}\!|\hat{y}_{1m}\, {\hat{y}}_{2n}^*|^2\,\big]^{1/2}$
	& $6.0	\times 10^{-4}$ \\[3ex]
		\multirow{2}{*}{$K_L^0 \to\pi^0  \nu \bar{\nu}$} &\multirow{2}{*}{ $<3.0\times 10^{-9}$} & 
	$\Im(\hat{y}_{1\ell}\, {\hat{y}}_{2\ell}^*)$
	& $[-1.1,\,1.2]\times 10^{-3}$ \\[0.5ex]
		&&$\big[\!\sum \limits_{m\not=n}\!|
		\hat{y}_{1m}\, {\hat{y}}_{2n}^* - \hat{y}_{2m}\, {\hat{y}}_{1n}^*
	|^2\,\big]^{1/2}$
& $1.1\times 10^{-3}$ \\[1.5ex]
		\hline
		\hline 
	\end{tabular}
	\caption{90\% C.L. bounds on leptoquark couplings from $K \to\pi \nu \bar{\nu}$ decays. The sign factor $s_{\mbox{\tiny LQ}}=+1$ for the leptoquarks $S_{1,3}$, while $s_{\mbox{\tiny LQ}}=-1$ for $\tilde{R}_2$.} \label{Table:KaonNu}
\end{table}

\begin{figure}[!h]
	\begin{center}
		\includegraphics[width=0.4\linewidth]{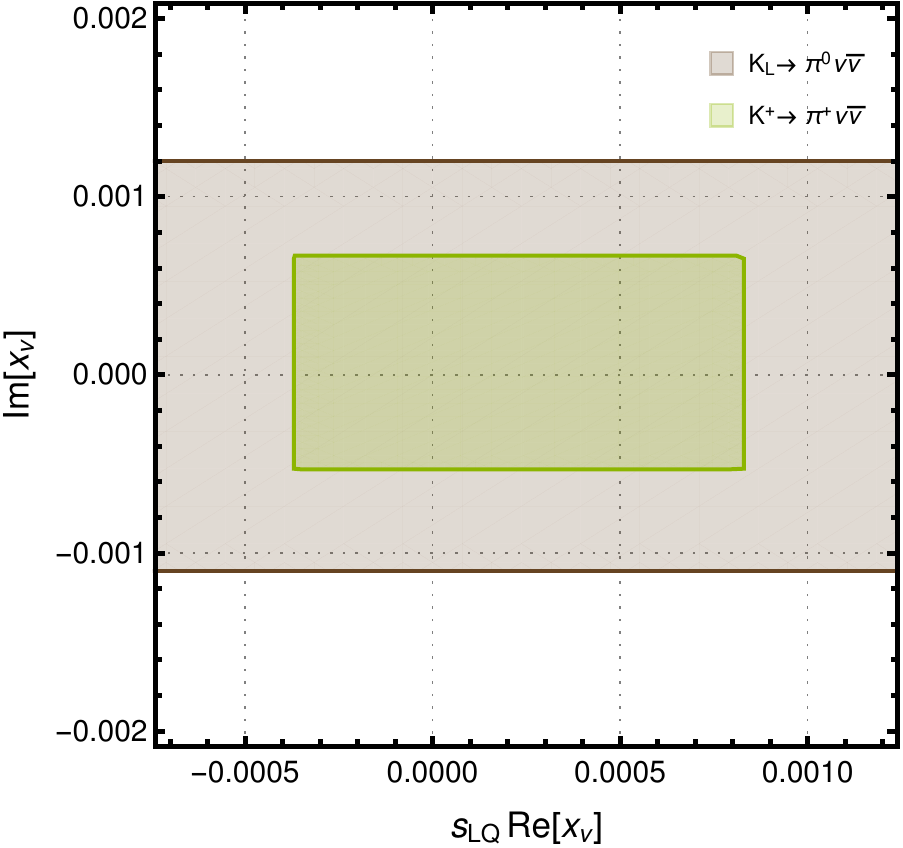}
		\caption{Allowed regions in the plane $(s_{\mbox{\tiny LQ}}\,\mathrm{Re}[x_\nu], \mathrm{Im}[x_\nu])$,
arising from $K\to\pi\nu\bar\nu$ decays.}\label{fig:KaonNu}
	\end{center}
\end{figure}

A constraint equation analogous to \eqn{eq:KpnunuCons} is obtained for $K_{L}^0\rightarrow\pi^{0}\nu\bar{\nu}$, but only the imaginary part of the relevant product of leptoquark couplings contributes to the decay into identical neutrino flavours. The extracted bounds, also shown in Table~\ref{Table:KaonNu}, are weaker than in the $K^+$ case because the current experimental sensitivity is not so good. 
The neutral and charged bounds, for identical neutrino flavours, are displayed in Fig.~\ref{fig:KaonNu}, where
\begin{equation}
x_\nu = \left(\frac{1~\mathrm{TeV}}{M_{\mathrm{LQ}}}\right)^2\times
\hat y_{1\ell}^{\phantom{*}} \hat y_{2\ell}^* 
\end{equation}
is the appropriate combination of leptoquark couplings.
Notice that the $\tilde S_1$ and $R_2$ leptoquarks do not generate contributions to these decay modes at tree level. A study on loop-induced effects in $K^+ \to\pi^+  \nu \bar{\nu}$ and $K_L^0 \to\pi^0  \nu \bar{\nu}$ can be found in Ref.~\cite{Fajfer:2018bfj},  for the $R_2$ and $S_3$ leptoquarks.

The KOTO collaboration has recently reported the observation of four events in the neutral decay mode~\cite{KOTO_kaon2019}, with an expected background
of only $0.05\pm 0.02$ events. Removing one of the events that is suspected to originate in underestimated upstream activity background, the quoted single event sensitivity of $6.9\times 10^{-10}$ would correspond to 
$\mathrm{BR}(K_L^0 \to\pi^0  \nu \bar{\nu}) \sim 2\times 10^{-9}$, well above the new Grossman-Nir limit~\cite{Grossman:1997sk} implied by the NA62 upper bound on $\mathrm{BR}(K^+ \to\pi^+  \nu \bar{\nu})$. This limit is valid under quite generic assumptions, provided the lepton flavour is conserved, and in the leptoquark case it can be directly inferred from Eqs.~(\ref{eq:Kppinunu}) and (\ref{eq:KLpinunu}). If there are only identical neutrino flavours in the final state, these two equations imply
\begin{equation}
\frac{\mathrm{BR}(K_L^0 \to\pi^0  \nu \bar{\nu})}{\mathrm{BR}(K^+ \to\pi^+  \nu \bar{\nu})}\, <\, \frac{\kappa_{\nu}^L \left( 1-\delta_\epsilon\right)}{\kappa_{\nu}^+ \left(  1+\Delta_{EM}\right)} \, =\, 4.2\, ,
\end{equation}
and, therefore, $\mathrm{BR}(K_L^0 \to\pi^0  \nu \bar{\nu}) < 7.8\times 10^{-10}$. The only way to increase this result and reach the KOTO signal would be through the decays into neutrinos with different flavours, $n\not= m$ in Eqs.~(\ref{eq:Kppinunu}) and (\ref{eq:KLpinunu}). Thus, a confirmation of the KOTO events would clearly indicate a violation of lepton flavour. Given their very preliminary status, we refrain from dwelling more on the physical meaning of these events. Some possible NP interpretations have been already considered in Ref.~\cite{Kitahara:2019lws}.

\subsection{$K^0-\bar{K^0}$ mixing}
\label{sec:KaonMix}

The leptoquarks contribute to kaon mixing via a box diagram mediated by leptons and leptoquarks similar to Fig.~\ref{dia:3l} (right panel) with the quark and lepton lines interchanged. The SM contribution to the off-diagonal element $M_{12}$ in the neutral kaon mass matrix is given by~\cite{Buras:1998raa}
\begin{align}
\label{eq:M12SM}
M_{12}^{\rm SM}=\; & \frac{\langle K^0|\mathcal{H}^{\rm SM}_{\Delta S=2}|\bar{K^0}\rangle}{2m_K} \nn \\
=\; & \frac{G_F^2 m_W^2}{12\pi^2}\, f_K^2 \hat{B}_K m_K \left[{\lambda_{c}^*}^2 \eta_{cc} S_0(z_c)+ {\lambda_t^*}^2 \eta_{tt} S_0(z_t)+ 2\lambda^*_c \lambda_t^* \eta_{ct} S_0(z_c,z_t)\right]\, ,
\end{align}
where $f_K$ is the kaon decay constant, $\hat{B}_K$ is the reduced bag parameter, the short-distance QCD effects are described through the correction factors $\eta_i$ and $S_0(z)$ are the Inami-Lim functions:
\begin{align}
S_0(z_c)&= z_c\, , \\
S_0(z_t)&= \frac{4z_t-11z_t^2+z_t^3}{4(1-z_t)^2}-\frac{3z_t^3 \,{\rm ln}z_t}{2(1-z_t)^3}\,, \\
S_0(z_c,z_t)&= z_c \left[{\rm ln}\frac{z_t}{z_c}- \frac{3z_t}{4(1-z_t)}- \frac{3z_t^2 {\rm ln}z_t}{4(1-z_t)^2}\right].
\end{align}
Here $z_{c,t}$ are defined as $z_{c,t}=m_{c,t}^2/m_W^2$.

A scalar leptoquark, with the interaction term $\left(\lambda_{ij}\bar{d}^i_{L,R}\,\ell_{R,L}^j +\lambda_{ij}^\prime \bar{d}^i_R\,\nu_L^j\right)\phi$,  gives rise to the following extra contribution to the $\Delta S=2$ Hamiltonian 
\begin{align}
\mathcal{H}^{\rm LQ}_{\Delta S=2}= \frac{1}{128 \pi^2 M_{\rm LQ}^2}\,
\left[\sum_{j=1}^{3}(\lambda_{1j}\lambda_{2j}^*+\lambda_{1j}^\prime\lambda_{2j}^{\prime*})\right]^2  (\bar{d}_{L,R} \gamma^\mu s_{L,R})(\bar{d}_{L,R} \gamma^\mu s_{L,R})\, .
\end{align}
Here we have neglected the contributions proportional to lepton masses which generate (pseudo)scalar operators $(\bar{d}_{R,L} s_{L,R})(\bar{d}_{R,L}s_{L,R})$.
Including the NP effect, the total dispersive matrix element can be written as
\begin{align}
\label{eq:M12Tot}
M_{12}=M_{12}^{\rm SM}+ \frac{1}{384 \pi^2 M_{\rm LQ}^2}\, f_K^2 \hat{B}_K m_K
\left[\sum_{j=1}^{3}(\lambda_{1j}\lambda_{2j}^*+\lambda_{1j}^\prime\lambda_{2j}^{\prime*})\right]^2.
\end{align}
The two observables $\Delta m_K$ and $\epsilon_K$ are related to $M_{12}$ as
\begin{eqnarray}
\label{eq:obs}
\Delta m_K \approx 2\, \Re\, M_{12}\, ,
\qquad\qquad\qquad
\epsilon_K \approx \kappa_\epsilon\,\frac{e^{i\pi/4}}{\sqrt{2}\Delta m_K}\, \Im\, M_{12}\, ,
\end{eqnarray}
where the phenomenological factor $\kappa_\epsilon\!=0.94\pm 0.02$ accounts for the estimated long-distance corrections to $\epsilon_K$~\cite{Buras:2010pza}.
The experimental measurements of these observables are~\cite{Tanabashi:2018oca}
\begin{eqnarray}
\Delta m_K = (3.484\pm 0.006)\times 10^{-15}\,{\rm GeV}\, ,
\qquad\qquad
|\epsilon_K|=(2.228\pm 0.011)\times 10^{-3}\, .
\end{eqnarray}
In the numerical analysis, we use estimates of various parameters in Eq.~\eqref{eq:M12SM} from~\cite{Aoki:2019cca,Cirigliano:2011ny} as
\begin{eqnarray}
\hat{B}_K\!=0.717\pm 0.024\, ,
\quad
\eta_{cc}\!=1.43\pm 0.23\, ,
\quad
\eta_{tt}\!=0.5765\pm 0.0065\, ,
\quad
\eta_{ct}\!=0.496\pm0.047\, .
\end{eqnarray}

In the SM the charm box diagram dominates the $CP$-conserving contribution to $M_{12}$ over the top loop effect, in spite of its large mass enhancement in the loop function, as the later is CKM suppressed. 
In addition, there are sizable long-distance contributions to $\Delta m_K$, which are difficult to quantify.
Hence we adopt a very conservative approach and allow the NP contributions alone (without the SM effect) to saturate the measured kaon mass difference. The resulting bounds are shown in Table~\ref{Table:mix}. 
However, $\Im\, M_{12}$ is well predicted in the SM and while using the expression for $\epsilon_K$ in Eq.~\eqref{eq:obs}, we take the measured value for $\Delta m_K$ and combine all theoretical and experimental uncertainties. To be explicit, we find the range for NP couplings for which $|\Im M_{12}|\approx (1.17\pm 0.03)\times 10^{-17}$ is satisfied. The results are depicted in Table~\ref{Table:mix}, where we separately show the contribution arising from charged leptons and neutrinos flowing in the loop. 

\begin{table}[t]
	\centering
	\renewcommand{\arraystretch}{1.3}
	\begin{tabular}{ c c c  }
		\hline \hline
		LQ &   Bound from $\Delta m_K$:  &  Range from $|\epsilon_K|$: 
		\\
		& $<7.1\!\times\! 10^{-4}\times \left(M_{\rm LQ}/{\rm TeV}\right)^2$  & $[-2.4,\,7.2]\!\times\! 10^{-6}\!\times \!\left(M_{\rm LQ}/{\rm TeV}\right)^2$
		\\[1ex] \hline \noalign{\vskip2pt}
		$R_2$ &  $\big|[\Re(\sum \limits_{i=1}^{3}y_{i2}y_{i1}^*)]^2\!-\![\Im(\sum \limits_{i=1}^{3}y_{i2}y_{i1}^*)]^2\big|$ & $\Re(\sum \limits_{i=1}^{3}y_{i2}y_{i1}^*)\,\Im(\sum \limits_{i=1}^{3}y_{i2}y_{i1}^*)$  
		 \\[2.2ex]
		$\tilde{R}_2,\tilde{S}_1,4\times S_3$   & $\big|[\Re(\sum \limits_{i=1}^{3}y_{1i}y_{2i}^*)]^2\!-\![\Im(\sum \limits_{i=1}^{3}y_{1i}y_{2i}^*)]^2\big|$ & $\Re(\sum \limits_{i=1}^{3}y_{1i}y_{2i}^*)\,\Im(\sum \limits_{i=1}^{3}y_{1i}y_{2i}^*)$
		\\[-1.5ex] (for $\ell$-loop) && 
		\\[.5ex] 
		$S_1,\tilde{R}_2,S_3$  	& $\big|[\Re(\sum \limits_{i=1}^{3}\hat{y}_{1i}\hat{y}_{2i}^*)]^2\!-\![\Im(\sum \limits_{i=1}^{3}\hat{y}_{1i}\hat{y}_{2i}^*)]^2\big|$ & $\Re(\sum \limits_{i=1}^{3}\hat{y}_{1i}\hat{y}_{2i}^*)\,\Im(\sum \limits_{i=1}^{3}\hat{y}_{1i}\hat{y}_{2i}^*)$ 
		\\[-1.5ex] (for $\nu$-loop) && 
		\\ \hline\hline 
	\end{tabular}
	\caption{Bounds on leptoquark couplings from neutral kaon mixing.}  \label{Table:mix}
\end{table}

\section{Summary and discussion}
\label{sec:summary}

In this work we have presented a detailed catalog of upper limits on scalar leptoquark interactions with SM fermions, arising from various decay modes of charged leptons and kaons. Compared to previous analyses, we attempted to be as much rigorous as possible to carefully include all contributions within the SM. We have first derived the most general low-energy four-fermion effective Lagrangian induced by tree-level scalar leptoquark exchange, and have worked out the particular values of its Wilson coefficients for the five possible types of leptoquarks.

We started with the decays of the tau lepton to pseudoscalar or vector meson states accompanied with a charged lepton. A few of these modes were examined in Ref.~\cite{Davidson:1993qk} (with the data available at that time), where the limits were obtained by comparing with the corresponding mode with neutrinos.  
The much stronger experimental upper bounds on these decays currently available imply substantially improved constraints on the leptoquark parameters from all channels, which
are presented in Tables~\ref{Table:meson} and \ref{Table:meson2}. 
The most stringent limits on scalar operators are obtained from $\tau \to \eta^\prime \ell$, while the $\tau \to \rho^0 \ell$ decay mode puts the strongest constraint on vector operators.

Transitions in purely leptonic systems can only be induced at the loop level.
Interestingly, the rare lepton-flavour-violating decay $\mu \to e \gamma$ is found to be immensely constraining for all scalar leptoquarks except $\tilde{R_2}$.
The analogous limits from $\tau \to e \gamma$ and $\tau \to \mu \gamma$
are also quite strong for the $R_2$ and $S_1$ leptoquarks, but much weaker for $\tilde S_1$ and $S_3$.
We have also shown that from the electric and magnetic dipole moment measurements of leptons, only the leptoquarks having interactions with both left- and right-handed quarks and leptons, i.e., $R_2$ and $S_1$, can be constrained. Essentially the top and/or charm quark going in the loop can enhance the rate for these two leptoquarks. 
The rare lepton decay $\ell\to\ell^\prime\ell^\prime\ell^{\prime\prime}$ can not compete with the corresponding radiative modes; however, taking into account all contributions from penguin and box diagrams, we have still derived constrained equations among different leptoquark couplings that must be satisfied. The expression for different lepton flavours in the final state has also been pointed out in this context.

Next we have investigated the rare decays of kaons, focusing on the very suppressed FCNC leptonic and semileptonic modes. We have derived the differential distributions of the $K\to\pi\ell^+\ell^-$ decays, taking into account all known effects within the SM. Owing to the strong suppression of the SM decay amplitudes, we have been able to derive useful limits on the leptoquarks couplings, even neglecting the SM contributions in some cases, e.g., $K^+ \to \pi^+ \ell^+\ell^-$ and $K_S^0 \to \pi^0 \ell^+\ell^-$. The decays $K_S^0 \to \pi^0 \ell^+\ell^-$ ($K_L^0 \to \ell^+\ell^-$) and $K_L^0 \to \pi^0 \ell^+\ell^-$ ($K_S^0 \to \ell^+\ell^-$) constrain the real and imaginary parts, respectively, of the same combination of leptoquark couplings, while $K^+ \to \pi^+ \ell^+\ell^-$ restricts its absolute value. 
The stronger constraints are extracted from the $K^0_L$ decays, owing to its long-lived nature that increases the sensitivity to the leptoquark contributions.
In addition to higher statistics and more accurate data, future improvements on these limits would require taking properly into account the interference between the SM and NP amplitudes, which in same cases it is currently hampered by poorly determined non-perturbative parameters.

We have also analyzed the strong constraints from $K\to\pi \nu \bar{\nu}$, taking into account the most recent limits from NA62 and KOTO. 
The recent four events observed by KOTO, which violate the Grossman-Nir limit, most probably originate in underestimated background/systematics. Nevertheless, we have pointed out that decay modes into neutrinos with different flavours could provide a possible explanation of the data in the leptoquark context.
For completeness, we have also compiled the constraints from
$K^0-\bar{K^0}$ mixing emerging from the one-loop leptoquark contributions.

\section*{Acknowledgments}
We thank Jorge Portol\'es and Avelino Vicente for useful discussions.
This work has been supported in part by the Spanish Government and ERDF funds from the EU Commission [Grant FPA2017-84445-P], the Generalitat Valenciana [Grant Prometeo/2017/053] and the Spanish Centro de Excelencia Severo Ochoa Programme [Grant SEV-2014-0398]. R.M. also acknowledges the support from the Alexander von Humboldt Foundation through a postdoctoral research fellowship.

\appendix

\section{Decay parameters}
\label{app:FF}

We adopt the usual definition of the decay constants as
\be 
\label{eq:para}
\langle 0 | \bar q^i\gamma^\mu\gamma_5 q^j |P_{ji}(k)\rangle\, =\, i f_P\, k^\mu\, ,
\qquad\qquad
\langle 0 |\bar{q}^i  \gamma^5 q^j| P_{ji}(k)\rangle =  \dsp\frac{-i\,  \,m_P^2 f_P}{(m_{q_i}+ m_{q_j})}
\, ,
\ee
%
%
\be 
\langle 0 | \bar q^i\gamma^\mu q^j |V_{ji}(k)\rangle\, =\, m_V f_V \,\varepsilon^\mu(k)\,,\qquad \langle 0 | \bar q^i\sigma^{\mu\nu} q^j |V_{ji}(k)\rangle\, =\, i f_V^\perp(\mu) \, [k^\mu\varepsilon^\nu(k)- k^\nu\varepsilon^\mu(k)]\, .
\ee
The values (in MeV) used in the numerical analysis are~\cite{Tanabashi:2018oca,Ali:1997nh,Jansen:2009hr,Mateu:2007tr,Cata:2009dq,Ball:2002ps}:
\begin{eqnarray}
\label{eq:decayconst}
&f_\pi= 132,~f_{K_S^0}=161,
~f_\rho=216,~f_\omega=195,~f_{K^{*0}}=f_{\overline{K}^{*0}}=214,~f_\phi=233,\nn \\ &f^\perp_\rho(1\gev)\approx f^\perp_\omega(1\gev)=160\,.
\end{eqnarray}
For the pseudoscalar mesons $\eta$ and $\eta^\prime$ we consider four different decay constants in the quark-flavour basis as
\begin{align}
f_\eta^q= f_q \cos \phi_q\, ,
\qquad\qquad
f_\eta^s= -f_s \sin \phi_s\, , 
\nn \\
f_{\eta^\prime}^q= f_q \sin \phi_q\, ,
\qquad\qquad
f_{\eta^\prime}^s= f_s \cos \phi_s\, .
\end{align}
Adopting the Leutwyler and Kaiser~\cite{Kaiser:1998ds} parametrization, the following values~\cite{Feldmann:1999uf} have been used in the analysis:
\begin{align}
f_q/f_\pi\simeq 1.08\, ,
\qquad\qquad
f_s/f_\pi\simeq 1.43\, ,
\qquad\qquad
\phi_q\simeq 44.8^o\, ,
\qquad\qquad
\phi_s\simeq 40.5^o\, .
\end{align}

\end{document}